\documentclass{amsart}

\usepackage{textcomp}
\usepackage{float}
\restylefloat{table}
\usepackage[utf8]{inputenc}
\usepackage[toc,page]{appendix}
\usepackage{amsmath,amsthm,amssymb,stmaryrd}
\usepackage{graphicx}
\usepackage[colorinlistoftodos,prependcaption]{todonotes}
\usepackage{xargs}
\usepackage{multicol}
\newcommandx{\unsure}[2][1=]{\todo[linecolor=blue,backgroundcolor=blue!25,bordercolor=blue,#1]{#2}}
\newmuskip\pFqmuskip

\newcommand*\pFq[6][8]{%
	\begingroup 
	\pFqmuskip=#1mu\relax
	\mathcode`\.=\string"8000
	\begingroup\lccode`\~=`\,
	\lowercase{\endgroup\let~}\pFqcomma
	{}_{\,#2}F_{\,#3}{\left[\genfrac..{0pt}{}{\,#4}{\,#5};#6\right]}%
	\endgroup
}
\newcommand{\pFqcomma}{\mskip\pFqmuskip}

\newtheorem{theorem}{Theorem}[section]

\newtheorem{lemma}{Lemma}[section]

\newcommand{\bbint}[2]{\ensuremath{\;\backslash\!\!\!\!\backslash\!\!\!\!\!\int_{#1}^{#2}}}

\begin{document}
\title[Generalized Stieltjes transform]{Finite-Part Integration of the Generalized Stieltjes Transform and its dominant asymptotic behavior for small values of the parameter Part I: Integer orders}

\author{Christian D. Tica}
\author{Eric A. Galapon}
\address{Theoretical Physics Group, National Institute of Physics, University of the Philippines, Diliman Quezon City, 1101 Philippines}
\email{eagalapon@up.edu.ph}
\date{\today}

\maketitle
\begin{abstract}
The paper addresses the exact evaluation of the generalized Stieltjes transform $S_{n}[f]=\int_0^{\infty} f(x) (\omega+x)^{-n}\mathrm{d}x$ of integral order $n=1,2, 3,\dots$ about $\omega =0$ from which the asymptotic behavior of $S_{n}[f]$ for small parameters $\omega$ is directly extracted. An attempt to evaluate the integral by expanding the integrand $(\omega+x)^{-n}$ about $\omega=0$ and then naively integrating the resulting infinite series term by term lead to an infinite series whose terms are divergent integrals. Assigning values to the divergent integrals, say, by analytic continuation or by Hadamard's finite part is known to reproduce only some of the correct terms of the expansion but completely misses out a group of terms. Here we evaluate explicitly the generalized Stieltjes transform by means of finite-part integration recently introduced in [E.A. Galapon, {\it Proc. Roy. Soc. A} {\bf 473}, 20160567 (2017)]. It is shown that, when $f(x)$ does not vanish or has zero of order $m$ at the origin such that $(n-m)\geq 1$, the dominant terms of $S_{n}[f]$ as $\omega\rightarrow 0$ come from contributions arising from the poles and branch points of the complex valued function $f(z) (\omega+z)^{-n}$. These dominant terms are precisely the terms missed out by naive term by term integration. Furthermore, it is demonstrated how finite-part integration leads to new series representations of special functions by exploiting their known Stieltjes integral representations. Finally, the application of finite part integration in obtaining asymptotic expansions of the effective diffusivity in the limit of high Peclet number, the Green-Kubo formula for the self-diffusion coefficient and the antisymmetric part of the diffusion tensor in the weak noise limit is discussed.
\end{abstract}

\section{Introduction}
The generalized Stieltjes transform of order $\lambda>0$ of a locally integrable function $f(x)$ in the interval $[0,\infty)$ is given by
\begin{equation}\label{generaltrans}
S_{\lambda}[f]=\int_0^{\infty} \frac{f(x)}{(\omega + x)^{\lambda}} \mathrm{d}x ,\,\,\,0<\omega<\infty
\end{equation}
provided the integral exists, at least, in the Riemann sense. The transform of order $\lambda=1$ is the standard Stieltjes transform,
\begin{equation}\label{standard}
S[f]=\int_0^{\infty} \frac{f(x)}{\omega +x} \, \mathrm{d}x ,
\end{equation}
whose classical properties was extensively studied in \cite{widder}. On the other hand, various properties of the generalized Stieltjes transform was studied separately in \cite{saxena,joshi,yurekli,schwarz,karp}. The application of Stieltjes transform and its generalization is not only limited to classical functions but has also been applied to transforms of distributions \cite{pathak,tekale,hayek}. Other generalizations of the Stieltjes transform \eqref{standard}, aside from equation \eqref{generaltrans}, are also known. These generalizations have kernels of transformations that reduce to the kernel of equation \eqref{generaltrans}, such as kernels in hypergeometric functions \cite{joshi,joshi2} and kernels involving powers of the variable $x$ \cite{lopez,hayek}. The generalized transform \eqref{generaltrans} has been a tool in function theory, such as in Weyl fractional calculus \cite{miana} and in representation theory of special functions \cite{ismail,karp2}.

In this paper we address the exact evaluation of equation \eqref{generaltrans} in the neighborhood of the origin, $\omega=0$, for positive integer values of $\lambda = n = 1,2,\dots$. The exact evaluation of the case for which $\lambda$ assumes non-integer values will be treated and presented separately in \cite{tica2}.  From the exact value of equation \eqref{generaltrans}, the correct asymptotic behavior of $S_{n}[f]$ as $\omega\rightarrow 0$ can then be obtained directly, a problem distinct from earlier asymptotic evaluations of the Stieltjes transform and its generalization in which the asymptotic expansion is sought in the opposite asymptotic regime $\omega\rightarrow\infty$ \cite{wong,distribution,mcwong,wong3,lopez}. An instinctive approach to the evaluation of $S_{n}[f]$ in the neighborhood of $\omega=0$ is to replace $(\omega+x)^{-n}$ with its binomial expansion about $\omega=0$, 
\begin{equation}\label{expansionaive}
\frac{1}{(\omega + x)^{n}} = \sum_{k=0}^{\infty} {-n \choose k} \frac{\omega^{k}}{x^{ k +  n}} ,
\end{equation}
in the integrand and then performing a term by term integration. This yields the infinite series of integrals
\begin{equation}
\sum_{k=0}^{\infty} {-n \choose k} \omega^{k} \int_0^{\infty} \frac{f(x)}{x^{ k +  n}}\mathrm{d}x .
\end{equation}
If the function $f(x)$ is analytic at the origin, the integrals are generally divergent due to the non-integrable singularity at the origin. One may attempt to give meaning to the divergent integrals by assigning values to them, say, by analytic continuation. This implies that $S_{n}[f]$ has a finite value at $\omega=0$; but when $n\geq 1$ and $f(0)\neq 0$ we expect divergence at the first term so that the integral is infinite when $\omega$ is zero. This indicates that a naive assignment of values to divergent integrals can lead to wrong results. 

It is known that term by term integration involving divergent integrals can lead to missing terms \cite{wong}. This was first pointed out and resolved by  McClure and Wong in the asymptotic evaluation of the Stieltjes transform \eqref{standard} as $\omega\rightarrow\infty$ \cite{mcwong}. There, they showed that interpreting the divergent integrals by analytic continuation leads to missing terms. The missing terms are recovered by interpreting the divergent integrals as functionals over some fundamental space of test functions so that the singular factors in the integrand are interpreted as distributions. However, despite of the absence of a counter example to the results of McClure and Wong (in fact we have recently utilized their distributional approach to exactify the Hankel transform \cite{kay}), it is desirable to clarify the intervening steps leading to their final results. In the intervening steps, divergent integrals arise and they are assigned values by analytic continuation. The use of analytic continuation there needs justification as it is already known that analytic continuation may leave out some terms.

Recently, one of us revisited the problem of missing terms arising from term by term integration involving divergent integrals in the standard Stieltjes transform \eqref{standard} in the neighborhood of $\omega=0$ without the use of distribution theory and analytic continuation \cite{galapon2}. There, the problem of missing terms is resolved by lifting the integration in the complex plane.  It is shown that the missing terms arise from the singularities of the complex valued function $f(z) (\omega + z)^{-1}$, with the divergent integrals arising from term by term integration interpreted as finite part integrals \cite{hadamard,fox,monegato,galapon1}. In particular, we had the result
\begin{equation}\label{original}
S[f]=\sum_{k=0}^{\infty} (-1)^k \omega^k \bbint{0}{\infty} \frac{f(x)}{x^{k+1}}\, \mathrm{d}x + \Delta_{\mathrm{sc}}(\omega)
\end{equation}
where the integral is the finite part of the divergent integral $\int_0^{\infty} f(x) x^{-k-1} \mathrm{d}x$ and $\Delta_{\mathrm{sc}}(\omega)$ is the contribution from the singularities, either from a pole or a branch point, of the function $f(z) (\omega+z)^{-1}$, in which $f(z)$ is the complexification of the function $f(x)$, obtained by replacing the real variable $x$ with the complex variable $z$ in $f$. The term $\Delta_{\mathrm{sc}}(\omega)$ is precisely the term missed out when performing naive term by term integration. 

The equality in equation \eqref{original} is not a mere asymptotic equality but an exact analytic equality. This leads to the use of divergent integrals, their finite parts in particular, in the analytical evaluation of a convergent integral. We have referred to this method of evaluating a convergent integral using the finite part of divergent integrals as finite-part integration \cite{galapon2}. In this paper we will consider the finite-part integration of the generalized Stieltjes integral \eqref{generaltrans}. We will obtain the general result
\begin{equation}
\int_0^{\infty} \frac{f(x)}{(\omega+x)^{n}}\, \mathrm{d}x = \sum_{k=0}^{\infty} {-n \choose k} \omega^{k} \bbint{0}{\infty} \frac{f(x)}{x^{ k + n}} \mathrm{d}x+ \Delta_{\mathrm{sc}}(\omega) .
\end{equation}
where $\Delta_{\mathrm{sc}}(\omega)$ constitutes the contributions coming from the singularities of $f(z)(\omega+z)^{-n}$ in the complex plane. Here we will refer to the first term as the naive contribution (or the naive term) as it arises from naive term by term integration, and the second term as the singular contribution (or the singular term) as it arises from the singularities of the integrand in the complex plane. We will obtain the explicit forms of $\Delta_{\mathrm{sc}}(\omega)$, and determine the dominant contributions in the asymptotic regime of arbitrarily small $\omega$. We will show that the singular contributions dominate the behavior of $S_{n}[f]$ as $\omega\rightarrow 0$ when $f(x)$ does not vanish or when it has a zero at the origin whose order does not sufficiently exceed the order of the Stieltjes transform. We will also demonstrate how finite-part integration leads to new representations of special functions. In particular, by exploiting known Stieltjes integral representations of the Gauss hypergeometric function and the Kummer function of the second kind, we will obtain new series representations of them. 

We accomplish our task here by way of the finite part-integration of the incomplete generalized Stieltjes transform given by
\begin{equation}\label{incomplete}
S_{n}^a[f]=\int_0^a \frac{f(x)}{(\omega+x)^{n}}\, \mathrm{d}x ,\;\; 0<a<\infty.
\end{equation}
The result for the generalized Stieltjes transform is obtained by means of the limit
\begin{equation}
\int_0^{\infty} \frac{f(x)}{(\omega+x)^{n}}\,\mathrm{d}x= \lim_{a\rightarrow \infty}\int_0^a \frac{f(x)}{(\omega+x)^{n}}\, \mathrm{d}x ,
\end{equation}
provided the limit exists in the standard sense. While the incomplete generalized Stieltjes transform is used to obtain the desired generalized Stieltjes transform, the incomplete transform (i.e. for some fixed upper limit of integration $a$) is important in itself. For example, special functions, such as the Gauss hypergeometric function, assume an incomplete generalized Stieltjes transform representation. Here we will limit ourselves to $f(x)$'s with entire complex extension $f(z)$. 
 
Finally the method developed here can be applied to the more general case of Stieltjes transform given by
\begin{equation}
\int_0^{\infty} \frac{f(x)}{(\omega^m + x^m)^n}\, \mathrm{d}x,\;\;\; m,n =1, 2, 3, \dots 
\end{equation}
when an evaluation of the transform is sought appropriate for small values of $\omega$. The special case $(m=2,n=1)$,
\begin{equation}\label{transform21}
\int_0^{\infty}\frac{f(x)}{\omega^2 + x^2}\, \mathrm{d}x,
\end{equation}
is relevant in several problems in physics. Possible applications of the transform defined by integral \eqref{transform21} include obtaining the asymptotic expansion of the effective diffusivity by passive advection in laminar and turbulent flows in the limit of high Peclet number \cite{avellada}, in the asymptotic expansion of the Green-Kubo formula for the self-diffusion coefficient and the antisymmetric part of the diffusion tensor
both in the weak noise limit \cite{pavliotis}. These physical quantities assume  generalized Stieltjes transform representations whose relevant integrals can be cast into the integral given by \eqref{transform21}. It happens that the desired asymptotic expansions correspond to the small $\omega$ evaluation of integral \eqref{transform21}; this makes finite part integration relevant in their investigations. It will suffice us to evaluate integral \eqref{transform21} for small $\omega$ using finite part integration. We will pursue elsewhere an in-depth study of these problems in the light of the progress reported here. 

The rest of the paper is organized as follows. In Section-\ref{finitepartintegration}, we discuss finite-part integration and outline how to implement finite-part integration of the generalized Stieltjes transform. In Section-\ref{seriesrep}, we develop general series representation of the finite part of divergent integrals arising from non-integrable singularity at the origin. In Section-\ref{integral} we consider the generalized Stieltjes transform of entire functions for integral orders. In Section-\ref{integralbranch}, we consider the Stieltjes transform of entire functions with branch point at the origin for integral orders. In Section-\ref{application}, we evaluate integral \eqref{transform21} using finite part integration, and discuss its possible applications to the high Peclet number expansion of the effective diffusivity in turbulent flows and to the weak noise limit of the Green-Kubo formula for the self-diffusion coefficient and the antisymmetric part of the diffusion tensor. In Section-\ref{conclusion}, we conclude. Analytical results are confirmed numerically using Mathematica\textsuperscript \textregistered  11.2 and Maple\textsuperscript \textregistered 18.00 on an Intel\textsuperscript \textregistered Core i7 processor with 8Gb of RAM.
 
\section{Finite-part integration}\label{finitepartintegration}
In this Section, we give an overview of finite-part integration and of how to implement it specifically to the evaluation of the incomplete generalized Stieltjes transform. Specific details are found later in Section-\ref{integral} and Section-\ref{integralbranch}. To perform finite-part integration on a given convergent integral, one generally has to proceed as follows, as initially outlined in \cite{galapon2}: 
 \begin{enumerate}
 \item  Determine the divergent integrals that arise after expanding the integrand of the given integral and performing a term by term integration; 
 
 \item obtain the finite parts of the divergent integrals; 
 
 \item obtain the complex contour integral representation of the finite parts; 
 
 \item represent the given integral as a complex contour integral using the same contour of integration as the finite part integrals; 
 
 \item perform the expansion of the integrand under the contour integral of the given (convergent) integral and proceed with the term by term integration; 
 
 \item recover the missing terms from the singularities of the integrand in the complex plane.
\end{enumerate}

We apply these steps in the evaluation of the integral (\ref{incomplete}). Since we wish to obtain an expansion that yields the asymptotic behavior as $\omega\rightarrow 0$, we use the binomial expansion of $(\omega+x)^{-n}$ about $\omega = 0$. Substituting this expansion back into the given integral leads to an infinite series of integrals
\begin{equation}\label{expansionxx}
\sum_{k=0}^{\infty} {-n \choose k} \omega^{k} \int_0^{a} \frac{f(x)}{x^{ k +  n}}\,\mathrm{d}x .
\end{equation}
The integrals are generally divergent when $f(x)$ is analytic at the origin due to the non-integrable singularity at the origin. We are then led to the identification of the divergent integrals arising from term by term integration. We accommodate for the possibility that $f(x)$ has a branch point at the origin so that we make the shift
\begin{equation}
f(x)\rightarrow x^{-\nu}f(x),\,\,\, 0\leq\nu<1,
\end{equation}
where $f(x)$ is now analytic at the origin and we recover equation \eqref{expansionxx} when $\nu=0$. Hence, in general, the relevant divergent integrals arising in equation \eqref{expansionxx} take the form
\begin{equation}\label{divergent}
\int_0^a \frac{f(x)}{x^{m+\nu}} \, \mathrm{d} x, \;\;\; 0\leq\nu<1, \; m=1, 2, \dots .
\end{equation}
This accomplishes the first step in the application of finite-part integration. 

Next is to obtain the finite part of these integrals. This is done by temporarily modifying the divergent integral to become convergent, followed by identifying the terms that diverge as the modified integral approaches the given divergent integral. There is no unique way of doing this but here the divergent integral is modified with the replacement of the offending non-integrable origin with some arbitrarily small $\epsilon$, $0<\epsilon<a$. Then the resulting convergent integral is grouped into two sets of terms
\begin{equation}\label{definitepart}
\int_{\epsilon}^{a} \frac{f(x)}{x^{m+\nu}} \mathrm{d}x = C_{\epsilon}+D_{\epsilon},
\end{equation}
where $C_{\epsilon}$ is the group of terms that possesses a finite limit as $\epsilon\rightarrow 0$, and $D_{\epsilon}$ is the group of terms that diverges in the same limit. The finite part of the divergent integral is obtained by dropping the diverging group of terms $D_{\epsilon}$, leaving only the limit of $C_{\epsilon}$ and assigning the limit as the value of the divergent integral,
\begin{equation}\label{finitepart}
\bbint{0}{a} \frac{f(x)}{x^{m+\nu}}\mathrm{d} x = \lim_{\epsilon\rightarrow 0} C_{\epsilon} .
\end{equation}
These foregoing steps outline the classical procedure by which finite part integrals are defined \cite{monegato}.

Next is to establish a contour integral representation of the finite part integral (\ref{finitepart}). The starting point for the construction of a complex contour integral representation is an equivalent form of the finite part integral which is given by
\begin{equation}\label{form2}
\bbint{0}{a} \frac{f(x)}{x^{m+\nu}} \mathrm{d}x = \lim_{\epsilon\rightarrow 0} \left[\int_{\epsilon}^a \frac{f(x)}{x^{m+\nu}} \mathrm{d}x - D_{\epsilon}\right]
\end{equation}
This follows from equation (\ref{definitepart}) and our chosen definition of the finite part. To obtain the contour integral representation, the function $f(x)$ is complexified by replacing the real variable $x$ with the complex variable $z$ to yield the complex valued function $f(z)$. We impose the minimum requirement that $f(z)$ is at least analytic in some neighborhood of the complex plane which takes the value $f(x)$ along the path of integration $[0,a]$ either from above or below the real axis. The complexified function $f(z)$ is then defined (uniquely) in the complex plane by analytic continuation. Now the contour integral representation of the finite part integral (\ref{form2}) depends on the analytic properties of $f(z)$, in particular, along the path of integration $[0,a]$. Either $f(z)$ is analytic along the path or it has at most a branch point there. The representation is obtained by using a contour $\mathrm{C}$ that starts at $a$ and goes around to enclose the segment $[0,a]$ back to the point $a$. Along this contour, the integral $\int_{\mathrm{C}} f(z) G(z) z^{-m-\nu} \mathrm{d}z$ is evaluated, where $G(z)$ is there to induce, when necessary, a branch cut along the path of integration. $G(z)$ is chosen such that when the contour $\mathrm{C}$ is continuously deformed to hug the segment $[0,a]$ the right hand side of equation (\ref{form2}) emerges in the limit. This is expected to lead to the desired complex contour integral representation of the finite part integral. 

\begin{figure}
	\includegraphics[scale=0.6]{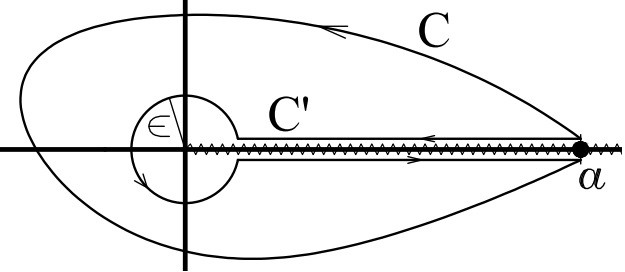}
	\caption{The contour of integration. The contour $\mathrm{C}$ does not enclose any pole of $f(z)$.}
	\label{tear}
\end{figure}

In \cite{galapon2}, we assumed that $f(z)$ is analytic in the segment $[0,a]$. Under this assumption, the contour integral representation will depend on the value of $\nu$. When $\nu=0$, the integrand $f(z) z^{-m}$ has a pole of order $m$ at the origin; when $\nu\neq 0$, it has a branch point there instead.  In \cite{galapon2}, we obtained the following complex contour integral representations of the finite part integral. 
\begin{theorem}\label{prop1}
	Let the complex extension, $f(z)$, of $f(x)$ be analytic in the interval $[0,a]$. If $f(0)\neq 0$, then 
\begin{equation}\label{result1}
	\bbint{0}{a}\frac{f(x)}{x^{m}}\mathrm{d}x=\frac{1}{2\pi i}\int_{\mathrm{C}} \frac{f(z)}{z^{m}} \left(\log z-\pi i\right)\mathrm{d}z, \;\; m = 1, 2 \dots 
\end{equation}
where $\log z$ is the complex logarithm whose branch cut is the positive real axis and $\mathrm{C}$ is the contour straddling the branch cut of $\log z$ starting from $a$ and ending at $a$ itself, as depicted in Figure-\ref{tear}.
\end{theorem}

\begin{theorem}
Let the complex extension, $f\left(z\right)$, of $f\left(x\right)$ be analytic in the interval $[0,a]$. If $f\left(0\right)\neq\,0$, then
\begin{align}\label{result2}
\bbint{0}{a}\,\frac{f\left(x\right)}{x^{m+\nu}}\,\mathrm{d}x=\frac{1}{\left(e^{-2\,\pi\,\nu\,i}-1\right)}\,\int_{C}\,\frac{f\left(z\right)}{z^{m+\nu}}\,\mathrm{d}z,\qquad m=1,2,\dots, 0<\nu<1,
\end{align}
where the branch of $z^{-\nu}$ is chosen such that it assumes positive values on top of the positive real axis and the contour $C$ is the path straddling the branch cut of $z^{-\nu}$ starting from and ending at $a$ itself, as depicted in Figure-\ref{tear}.
\end{theorem}
\noindent Observe that in this representation, the finite part integral is the value of an absolutely convergent integral similar to what we have earlier established for the Cauchy principal value and the Fox principal value \cite{fox,galapon1}. This lifts the rather vague meaning of the finite part into a well-defined convergent integral.

\begin{figure}
	\includegraphics[scale=0.5]{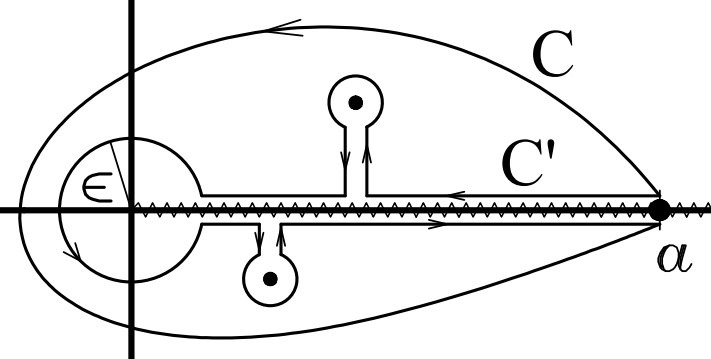}
	\caption{The contour of integration in the contour integral representation of a real integral. The contour $\mathrm{C}$ encloses some or all the poles of the integrand.}
	\label{fig:boat1}
\end{figure}

Once the contour integral representation of the relevant finite part integrals has been obtained, we obtain a similar contour integral representation of the given convergent integral using the same contour of integration.  For $f(z)$'s that are analytic along $[0,a]$, we have the following contour integral representations.
\begin{lemma} \label{lemma1}
Let $g(x)$ be integrable in the interval $[0,a]$ and that its complex extension $g(z)$ is analytic in a region containing the interval $[0,a]$ in its interior and holomorphic elsewhere. Then
\begin{equation}\label{contourep1}
	\int_0^a g(x) \, \mathrm{d}x = \frac{1}{2\pi i} \int_{\mathrm{C}} g(z) \log z \, \mathrm{d}z - \sum_k \mathrm{Res}\left[\log z \, g(z)\right]_{z_k},
\end{equation}
where the branch cut of $\log z$ is chosen to be the positive real axis and $\mathrm{C}$ is the contour straddling the branch cut of $\log z$ starting from $a$ and ending at $a$ itself, as depicted in Figure-2, and the $z_k$'s are the poles of $g(z)$ enclosed by $\mathrm{C}$, with no pole of $g(z)$ lying along $\mathrm{C}$.
\end{lemma}
\begin{lemma} \label{lemma0}
	Let $g(x)$ be integrable in the interval $[0,a]$ and that its complex extension $g(z)$ is analytic in a region containing the interval $[0,a]$ in its interior and holomorphic elsewhere. Then
	\begin{equation}\label{contourep2}
	\int_0^a x^{-\nu}\,g(x) \, \mathrm{d}x = \frac{1}{\left(e^{-2\,\pi\,\nu\,i}-1\right)} \int_{\mathrm{C}} z^{-\nu}g(z)\, \mathrm{d}z - \frac{2\,\pi\,i}{\left(e^{-2\,\pi\,\nu\,i}-1\right)}\sum_k \mathrm{Res}\left[z^{-\nu} \, g(z)\right]_{z_k},
	\end{equation}
where the branch cut of $z^{-\nu}$ is chosen to be the positive real axis and $\mathrm{C}$ is the contour straddling the branch cut of $z^{\nu}$ starting from $a$ and ending at $a$ itself, as depicted in Figure-2, and the $z_k$'s are the poles of $g(z)$ enclosed by $\mathrm{C}$, with no pole of $g(z)$ lying along $\mathrm{C}$.
\end{lemma}

Specific implementation of the contour integral representation of the incomplete generalized Stieltjes transform depends on the order $\lambda$ of the transformation, and the analytic properties of $f(z)$, and the value of the parameter $\nu$. For integral orders $\lambda=n=1, 2, \dots$, $\nu=0$ and entire $f(z)$, the Stieltjes transform will have to take the representation given by equation \eqref{contourep1}; this is implemented in Section-\ref{integral}. On the other hand, for integral orders $\lambda=n$, $\nu\neq 0$ and entire $f(z)$, the Stieltjes transform will assume the representation given by equation \eqref{contourep2}; this is implemented in Section-\ref{integralbranch}. For non-integral order $\lambda\neq n$, a contour integral representation other than equations \eqref{contourep1} and \eqref{contourep2} will have to be devised in \cite{tica2}.

Comparing the contour integral representations of the finite part integrals in the Theorems and the first terms in the Lemmas above, our intention of representing the given convergent integral as a contour integral using the same contour as in the finite part integral is apparent. The reason is that when the complexified version of the expansion \eqref{expansionaive}, given by
\begin{equation}\label{expansion2}
\frac{1}{(\omega + z)^{n}} = \sum_{k=0}^{\infty} {-n \choose k} \frac{\omega^{k}}{z^{ k + n}} ,
\end{equation}
is substituted back into the contour integral representation of the incomplete generalized Stieltjes transform and term by term integration is implemented, we will be able to identify the contour integrals that appear as finite parts of the divergent integrals that arise from naive term by term integration. 

Not only that the contour integral representation allows unique identification of the divergent integrals as finite part integrals, it is responsible in picking up the poles and branch points of the integrand $f(z)(\omega+z)^{-n}$ that are the origins of the missing terms in the naive application of term by term integration of the expansion \eqref{expansionxx}. The missing terms are precisely the residue terms in equations \eqref{contourep1} and \eqref{contourep2}.

\label{remrk}\subsubsection*{Remark 1} From the definition of the finite part integral given by equation \eqref{finitepart}, it is evident that, when  the divergent part $D_{\epsilon}$ vanishes, the finite part integral is just the (Riemann) improper integral. Also the complex contour integral representations of the finite parts given by equations \eqref{result1} and \eqref{result2} reduce to the regular (Riemann) integrals of the integrands when the improper integrals exist. For this reason, we can always replace the regular integral $\int_0^a$ with the finite part integral $\bbint{0}{a}$ without possible confusion for the two values coincide when the former exists. In short, the finite part integral of a convergent integral is just the value of the convergent integral itself.

\subsubsection*{Remark 2} In general the finite part integral is denoted by $\mathrm{FPI}\!\int$ but here we choose the notation $\bbint{}{}$, in keeping with the notation in \cite{galapon2}. The reason is that the sufficient condition for the existence of the finite part is differentiability of the the function $f(x)$ up to some finite order, i.e., it is not necessarily infinitely differentiable. However, in order for the complex contour integral representations \eqref{result1} and \eqref{result2} to hold it is necessary to impose the condition that the complexification of $f(x)$, $f(z)$, is analytic in some region containing the segment $[0,a]$. This implies that $f(x)$ is necessarily infinitely differentiable in $[0,a]$; the notation $\bbint{}{}$ serves to indicate that $f(x)$ has this property.

\section{Series representation of finite part integrals arising from non-integrable singularity at the origin}\label{seriesrep}
In this Section, we obtain the series representation of the finite part of the divergent integral (\ref{divergent}) to facilitate its calculation in the rest of the paper. We will assume that the complexified function $f(z)$ is entire so that it posseses a power series expansion with an infinite radius of convergence,
\begin{equation}
f(z)=\sum_{k=0}^{\infty} c_k z^k ,
\end{equation}
where the $c_k$'s are constants. We will derive the representation in two ways. First, by means of the definition of the finite part integral; and then by means of the complex contour integral representation of the finite part of the divergent integral. This provides an explicit demonstration of the equivalence of the classical definition and the contour integral representation of the finite part integral which is not immediately apparent. 

\subsection{Case $\nu=0$}
We now derive the series representation of the finite part integral $\bbint{0}{a} x^{-m} f(x) \mathrm{d}x$ using the classical definition of the finite part of a divergent integral given by equation \eqref{finitepart}. For some positive $\epsilon<a$, we identify the convergent and divergent parts of the (convergent) integral $\int_{\epsilon}^a f(x) x^{-m}\mathrm{d}x$ as $\epsilon\rightarrow 0$. We replace $f\left(x\right)$ with its Taylor series expansion, $   f\left(x\right) =  \sum_{k=0}^{\infty}\,c_{k}\,x^{k}$, in the integral. Since the limits of integration are well within the radius of convergence of the series, a term-wise integration is warranted. Splitting the summation into three parts: $0\le k\le m-2$, $k=m-1$, and $m\le k<\infty$, and performing the integration we obtain 
	\begin{eqnarray}\nonumber
  \int_{\epsilon}^{a}\,\frac{f\left(x\right)}{x^{m}}\,\mathrm{d}x&=&\sum_{k=0}^{m-2}\,\frac{c_{k}}{\left(k-m+1\right)}\,\left(\frac{1}{a^{m-k-1}}-\frac{1}{\epsilon^{m-k-1}}\right) + c_{m-1}\,\left(\ln\,a-\ln\,\epsilon\right)\\
    &&+ \sum_{k=m}^{\infty}\,\frac{c_{k}}{\left(k-m+1\right)}\,\left(a^{k-m+1}-\epsilon^{k-m+1}\right)
	\end{eqnarray}
Taking the limit as $ \epsilon \to 0$ in the equation above, we identify the converging and diverging terms as follows 
    \begin{align}
    C_{\epsilon} = c_{m-1} \ln a + \sum_{k=m}^{\infty} \frac{c_k}{(k-m+1)}\,\left(a^{k-m+1}-\epsilon^{k-m+1}\right)-\sum_{k=0}^{m-2}\,\frac{c_{k}}{\left(m-k-1\right)}\,\frac{1}{a^{m-k-1}},
    \end{align}
    \begin{align}
    D_{\epsilon} = -c_{m-1}\,\ln\,\epsilon + \sum_{k=0}^{m-2}\,\frac{c_{k}}{\left(m-k-1\right)}\,\frac{1}{\epsilon^{m-k-1}} .
    \end{align}
    
 Dropping $D_{\epsilon}$ altogether and assigning the value $\lim_{\epsilon\to 0}\,C_{\epsilon}$ to the divergent integral, we obtain the series representation of the finite part integral
\begin{equation}\label{polefinitepart}
    \bbint{0}{a} \frac{f(x)}{x^{m}} \, \mathrm{d}x = c_{m-1} \ln a - \sum_{k=0}^{m-2} \frac{c_k}{(m-k-1)} \frac{1}{a^{m-k-1}} +\sum_{k=m}^{\infty} \frac{c_k\,a^{k-m+1}}{(k-m+1)} .
\end{equation}
Moreover, taking the limit as $a \to\,\infty$, we obtain the finite part integral
\begin{equation}\label{polelimit}
\bbint{0}{\infty} \frac{f(x)}{x^{m}} \, \mathrm{d}x = \lim_{a\rightarrow\infty} \left[c_{m-1} \ln a + \sum_{k=m}^{\infty} \frac{c_k\,a^{k-m+1}}{(k-m+1)}\right],
\end{equation}
provided the limit exists.
    
We now recover the result \eqref{polefinitepart} using the contour integral representation of the finite part integral \eqref{result1}. We deform the contour of integration $C$ in Figure-\ref{tear} into a circle of radius $a$ centered at the origin; since $f(z)$ is entire, the value of the original contour integral is equal to the value of the integral along the circle. With the parametrization $z=a e^{i\theta}$ along the circular path of integration, the contour integral representation assumes the form
\begin{equation}\label{qik}
\bbint{0}{a} \frac{f(x)}{x^m} \mathrm{d}x = \frac{i}{a^{m-1}} \frac{1}{2\pi i} \int_0^{2\pi} f(a e^{i \theta}) \left[\ln a + i (\theta-\pi)\right] e^{i (1-m)\theta} \mathrm{d}\theta
\end{equation}
The integrals in the right hand side are evaluated by replacing $f(a e^{i\theta})$ with its expansion
\begin{equation}
f(a e^{i\theta}) = \sum_{k=0}^{\infty} c_k a^k e^{i k \theta}
\end{equation}
and then integrating term by term, which we can do again because the series converges uniformly along the path of integration.

The pair of integrals in \eqref{qik} can be evaluated using the following integral identities 
\begin{equation}
	\int_0^{2\pi} \mathrm{e}^{-i (n-k)\theta}\mathrm{d}\theta=\left\{\begin{array}{cc}
	0 &, n\neq k\\
	2\pi &, n=k
	\end{array}
	\right. ,
	\end{equation}
	\begin{equation}
	\int_0^{2\pi} \mathrm{e}^{-i (n-k)\theta}\theta \mathrm{d}\theta=\left\{\begin{array}{cc}
	\frac{2\pi i}{(n-k)} &, n\neq k\\
	2\pi^2 &, n=k
	\end{array}
	\right. .
	\end{equation}
 By application of these identities, we obtain the following integrals
 \begin{equation}
 \int_0^{2\pi} f(a e^{i\theta}) e^{i (1-m)\theta} \mathrm{d}\theta = 2\pi c_{m-1} a^{m-1}
 \end{equation}
 \begin{equation}
 \int_0^{2\pi} f(a e^{i\theta}) e^{i (1-m)\theta} \theta \,\mathrm{d}\theta = 2\pi i\sum_{k=0}^{m-2} \frac{c_k}{(m-k-1)} a^k + 2\pi^2 c_{m-1} + \sum_{k=m}^{\infty}\frac{c_k}{(m-k-1)} a^k
 \end{equation}
 Substituting them back in equation \eqref{qik} reproduces the finite part integral \eqref{polefinitepart}.

We now establish under what condition the limit $a\rightarrow\infty$ exists in equation \eqref{polelimit}. We use the contour integral representation \eqref{result1} of the finite part integral. We deform the contour $C$ in Figure-\ref{tear} into the key-hole contour $C'$. The circular part of the contour has a radius $\epsilon<a$. We then have
\begin{equation}
\bbint{0}{a} \frac{f(x)}{x^m}\, \mathrm{d}x = \frac{1}{2\pi i} \int_{\mathrm{C}_{\epsilon}} \frac{f(z)}{z^m} (\log z - i\pi)\, \mathrm{d}z + \int_{\epsilon}^{a} \frac{f(x)}{x^m} \, \mathrm{d}x .
\end{equation}
The first term is independent of the upper limit of integration $a$ so that only the second term is relevant in the limit as $a$ becomes arbitrarily large. Clearly the limit in equation \eqref{polelimit} exists provided $f(x)x^{-m}$ is integrable at infinity. Then we have proved the following series and limit representation of the finite part integral.

\begin{theorem}\label{theoremrep}
Let  $f(x)$ have an entire complex extension $f(z)$ such that it admits the expansion
\begin{equation}
f(z)=\sum_{k=0}^{\infty} c_k z^k .
\end{equation}
Then
\begin{equation}\label{poles}
\bbint{0}{a} \frac{f(x)}{x^{m}} \, \mathrm{d}x = c_{m-1} \ln a - \sum_{k=0}^{m-2} \frac{c_k}{(m-k-1)} \frac{1}{a^{m-k-1}} +\sum_{k=m}^{\infty} \frac{c_k a^{k-m+1}}{(k-m+1)}, \; m=1, 2, \dots 
\end{equation}
in which an empty sum is zero. Moreover,
\begin{equation}\label{FPI1}
\bbint{0}{\infty} \frac{f(x)}{x^{m}} \, \mathrm{d}x = \lim_{a\rightarrow\infty} \left[c_{m-1} \ln a + \sum_{k=m}^{\infty} \frac{c_k a^{k-m+1}}{(k-m+1)}\right],
\end{equation}
provided  $f(x) x^{-m}$ is integrable at infinity or $f(x) x^{-m}=o(x^{-1})$ as $x\rightarrow\infty$.
\end{theorem}

\subsubsection{Example} We apply Theorem-\ref{theoremrep} to determine the finite part of the divergent integral $\int_0^{\infty} x^{-m} e^{- b x}\, \mathrm{d}x$ for positive integer $m$ and $b>0$. We identify $f\left(x\right) = e^{-b\,x}$ which has an entire complex extension with the expansion coefficients $c_k=(-b)^k/k!$. Substituting the coefficients $c_{k}$ back into equation \eqref{FPI1}, we obtain the finite part integral in limit form
\begin{equation}\label{bababa}
    \bbint{0}{\infty}\,\frac{e^{-bx}}{x^{m}}\,\mathrm{d}x = \lim_{a\rightarrow\infty}\left(\frac{\left(-1\right)^{m-1}\,b^{m-1}}{\left(m-1\right)!}\,\ln\,a+\sum_{k=m}^{\infty}\frac{\left(-1\right)^{k}\,b^{k}\,a^{k-m+1}}{k!\,\left(k-m+1\right)}\,\right).
\end{equation}
To facilitate the calculation of the limit, we write the infinite sum in the second term as a hypergeometric function, 
\begin{equation}  \sum_{k=m}^{\infty}\frac{\left(-1\right)^{k}\,b^{k}\,a^{k-m+1}}{k!\,\left(k-m+1\right)} = \frac{\left(-1\right)^{m}\,a\,b^{m}}{m\,!}{_2F_2}\left(1,\,1;\,m+1,\,2;\,-a\,b\,\right).
\end{equation}
    
We then make use of the asymptotic expansion of the hypergeometric function ${_2F_2}$ for the case of a double pole for large arguments. The relevant expansion is given by \cite{doublepole2F2},
\begin{eqnarray}\label{pakoo}
 &&\hspace{-4mm}{_2F_2}\left(a_1,\,a_1;\,b_1,\,b_2;\,z\,\right) = \frac{\Gamma\left(b_1\right)\,\Gamma\left(b_2\right)}{\Gamma\left(a_1\right)^{2}}\,e^{z}\,\left(1+\mathcal{O}\left(z^{-1}\right)\,\right)\,z^{2\,a_{1}-b_{1}-b_{2}}\\\nonumber    &&\hspace{2mm}+\,\,\frac{\Gamma\left(b_1\right)\,\Gamma\left(b_2\right)\,\left(-z\right)^{-a_1}}{\Gamma\left(a_1\right)\,\Gamma\left(b_1-a_1\right)\,\Gamma\left(b_2-a_1\right)}\,\left[\,\log\left(-z\right)\,\left(1+\mathcal{O}\left(z^{-1}\right)\right)\right.\\\nonumber
&&  \hspace{2mm}  \left.-\left(\,\psi\left(b_1-a_1\right)+\psi\left(b_2-a_1\right)+\psi\left(a_1\right)+2\gamma\,\right)\,\left(1+\mathcal{O}\left(z^{-1}\right)\right)\right],\,\,|z|\to\,\infty ,
\end{eqnarray}
where $\psi\left(z\right)$ is the logarithmic derivative of the gamma function and $\gamma = -\psi\left(1\right)$ is the Euler constant. By inspection, only the second term in the expansion \eqref{pakoo} contributes in the limit as  $a\rightarrow\infty$. Substituting the leading contribution of ${_2F_2}$ back into equation \eqref{bababa} and taking the limit, we obtain the finite part integral
\begin{equation}\label{expo1}
    \bbint{0}{\infty}\,\frac{e^{-bx}}{x^{m}}\,\mathrm{d}x = \frac{\left(-1\right)^{m}\,b^{m-1}}{\left(m-1\right)!}\left(\ln\,b-\psi\left(m\right)\right) .
    \end{equation}

Observe that naive application of change variable by substitution does not generally hold for the finite part integral. The finite part integral \eqref{expo1} demonstrates this. When $b=1$ the logarithmic term vanishes giving
\begin{equation}\label{kiw2}
\bbint{0}{\infty}\frac{e^{-x}}{x^m}\mathrm{d}x=-\frac{\left(-1\right)^{m}}{\left(m-1\right)!}\psi\left(m\right).
\end{equation}
This result was earlier obtained in \cite{galapon2}. An attempt to evaluate the right hand side of equation \eqref{bababa} by changing variable $x\rightarrow x/b$ and then using the specific value \eqref{kiw2} leads to the finite part
\begin{align}\label{kiw3}
\bbint{0}{\infty}\frac{e^{-b x}}{x^m}\mathrm{d}x=-\frac{\left(-1\right)^{m} b^{m-1}}{\left(m-1\right)!}\psi\left(m\right).
\end{align}
Comparing equations \eqref{expo1} and \eqref{kiw3} we see that a naive change of variable misses out the logarithmic term.

\subsection{Case $\nu\neq 0$} As in the previous case, we can either use the classical definition of the finite part integral or its contour integral representation to obtain the series representation of the finite part integral. The result is summarized by the following statement.
\begin{theorem}\label{fpi_nu_not_0}
Let  $f(x)$ have an entire complex extension $f(z)$ such that it admits the expansion
\begin{equation}
f(z)=\sum_{k=0}^{\infty} c_k z^k .
\end{equation}
Then
\begin{equation}\label{branchpoint}
\bbint{0}{a} \frac{f(x)}{x^{m+\nu}} \, \mathrm{d}x = \sum_{k=0}^{\infty} \frac{c_k a^{k+1-m-\nu}}{(k+1-m-\nu)}, \; 0<\nu<1, \; m=1, 2, \dots 
\end{equation}
Moreover,
\begin{equation}\label{infinity_si_a}
\bbint{0}{\infty} \frac{f(x)}{x^{m+\nu}} \, \mathrm{d}x = \lim_{a\rightarrow\infty} \sum_{k=m}^{\infty} \frac{c_k a^{k+1-m-\nu}}{(k+1-m-\nu)},
\end{equation}
provided $f(x) x^{-m-\nu}$ is integrable at infinity or $f(x) x^{-m-\nu}=o(x^{-1})$ as $x\rightarrow\infty$.
\end{theorem}
\subsubsection{Example} We apply  Theorem-\ref{fpi_nu_not_0} to obtain the finite part of the divergent integral $\int_0^{\infty} x^{-m-\nu} e^{-b x}\, \mathrm{d}x$ for the parameters of the theorem and for $b>0$. Substituting the coefficients $c_k=(-b)^k/k!$ back in equation \eqref{infinity_si_a}, we obtain the finite part in limit form
\begin{equation}
	\bbint{0}{\infty}\,\frac{e^{-b\,x}}{x^{m+\nu}}\,\mathrm{d}x = \lim_{a\to\infty}\,\sum_{k=m}^{\infty}\,\frac{\left(-1\right)^{k}\,b^{k}\,a^{k+1-m-\nu}}{k!\,\left(k+1-m-\nu\right)} .
\end{equation}
Again to facilitate the calculation of the limit, we sum the series in terms of a hypergeometric function,
\begin{equation}
\sum_{k=m}^{\infty}\,\frac{\left(-1\right)^{k}\,b^{k}\,a^{k+1-m-\nu}}{k!\,\left(k+1-m-\nu\right)} = \frac{\left(-1\right)^{m+1}\,a^{1-\nu}\,b^{m}}{m!\,\left(\nu-1\right)}\,{_2F_2}\,\left(1,\,1-\nu;\,m+1,\,2-\nu;\,-a\,b\right) .
\end{equation}

We then make use of its asymptotic expansion for large arguments for the case of simple poles \cite{simplepole2F2}, 
\begin{eqnarray}	{_2F_2}\left(a_1,\,a_{2};\,b_{1},\,b_{2};\,z\right)&=& \frac{\Gamma\left(b_1\right)\,\Gamma\left(b_2\right)\,\Gamma\left(a_2-a_1\right)}{\Gamma\left(a_2\right)\,\Gamma\left(b_1-a_1\right)\,\Gamma\left(b_2-a_1\right)}\,\left(-z\right)^{-a_1}\,\left(1+\mathcal{O}\left(z^{-1}\right)\right)\nonumber\\
&&    +\,\,\frac{\Gamma\left(b_1\right)\,\Gamma\left(b_2\right)\,\Gamma\left(a_1-a_2\right)}{\Gamma\left(a_1\right)\,\Gamma\left(b_1-a_2\right)\,\Gamma\left(b_2-a_2\right)}\,\left(-z\right)^{-a_2}\,\left(1+\mathcal{O}\left(z^{-1}\right)\right)\nonumber\\
&&    +\,\,\frac{\Gamma\left(b_1\right)\,\Gamma\left(b_2\right)}{\Gamma\left(a_1\right)\,\Gamma\left(a_2\right)}\,e^{z}\,z^{a_1+a_2-b_1-b_2}\,\left(1+\mathcal{O}\left(z^{-1}\right)\right)
\end{eqnarray}
By inspection, the second term dominates all the other terms for the given parameter for arbitrary large $z$. Then in the limit as $a\rightarrow\infty$, we obtain the finite part integral 
\begin{equation}\label{expofinite}
	\bbint{0}{\infty}\,\frac{e^{-bx}}{x^{m+\nu}}\,\mathrm{d}x = \frac{\left(-1\right)^{m}\,b^{m+\nu-1}\,\pi}{\sin\left(\pi\,\nu\right)\,\Gamma\left(m+\nu\right)} .
\end{equation}

Observe that substitution works this time. The value of the finite part integral \eqref{expofinite} at $b\neq 1$ can be obtained from its value at $b=1$ by mere substitution. This is in contrast with the earlier case where the value at $b=1$ does not determine the value at other values of $b$. This shows that integration by substitution in real integration is not a property enjoyed by finite part integration. This should not come as a surprise. The reason is that the finite part integral is in fact not an integral in the real line but a contour integral in the complex plane where singularities enclosed by the contour of integration make contributions. It is these contributions that mere substitution does not capture. The correct way to perform integration by substitution in finite part integration is by means of the complex contour integral of a given finite part integral.  

\subsection{Polynomials} Polynomials are special class of entire functions, so that the above results apply. Here we consider the finite part integrals
\begin{equation}\label{koly}
\bbint{0}{a} \frac{P_{[r]}^{[s]}(x)}{x^{m+\nu}}\, \mathrm{d}x ,
\end{equation}
for $0\leq \nu<1$, where $P_{[r]}^{[s]}(x)$ is the polynomial 
\begin{equation}
P_{[r]}^{[s]}(x)=\sum_{k=r}^s a_k x^k
\end{equation}
where the $a_k$'s are constants independent of $x$ and $a_k\neq 0$ for $k=r,s$. That is $P_{[r]}^{[s]}(x)$ is a polynomial of order $s$ that has zero of order $r$ at the origin.

For pole singularities, $\nu=0$, equation \eqref{poles} has several cases depending on the relative values of $m$ and the order $s$ of the polynomial. By inspection, we have the following results:
\begin{equation}\label{koly}
\bbint{0}{a} \frac{P_{[r]}^{[s]}(x)}{x^{m}}\, \mathrm{d}x = \int_{0}^{a} \frac{P_{[r]}^{[s]}(x)}{x^{m}}\, \mathrm{d}x=\sum_{k=r}^s \frac{a_k a^{k-m+1}}{(k-m+1)} , \;\; m<r,
\end{equation}
\begin{equation}\label{integer_polynomial}
\bbint{0}{a} \frac{P_{[r]}^{[s]}(x)}{x^{m}}\, \mathrm{d}x = -\sum_{k=r}^s \frac{a_k}{(m-k+1) a^{m-k+1}}, \;\; m>s ,
\end{equation}
\begin{eqnarray}
&&\bbint{0}{a} \frac{P_{[r]}^{[s]}(x)}{x^{m}}\, \mathrm{d}x = -\sum_{k=r}^{m-2} \frac{a_k}{(m-k+1) a^{m-k+1}} + a_{m-1} \ln a + \sum_{k=m}^s \frac{a_k a^{k-m+1}}{(k-m+1)},\label{bebebe} \\
&&\hspace{80mm}r+1\leq m\leq s+1 \nonumber ,
\end{eqnarray}
where an empty sum in equation \eqref{bebebe} is assigned the value zero. In equation \eqref{koly}, the finite part is equal to the regular integral as the integral exists, in accordance with Remark 1 in Section-\ref{finitepartintegration}. 

On the other hand, in the presence of a branch point singularity, for $\nu\neq 0$, equation \eqref{branchpoint} reduces to the following single result:
\begin{equation}\label{polynomialbranch}
\bbint{0}{a} \frac{P_{[r]}^{[s]}\left(x\right)}{x^{m+\nu}} \, \mathrm{d}x = \sum_{k=r}^{s} \frac{a_k\,a^{k+1-m-\nu}}{(k+1-m-\nu)}.
\end{equation}

\section{Generalized Stieltjes transform of integral order of entire functions}\label{integral}

In this section, we evaluate by finite-part integration the incomplete generalized Stieltjes transform of integral order $n$ of a function $f(x)$, 
\begin{equation}
S_{n}^a[f]=\int_0^a \frac{f(x)}{(\omega+x)^n}\, \mathrm{d}x, \;\; 0<\omega<\infty, \;\; n=1, 2, \dots .
\end{equation}
To proceed with the finite-part integration, we have to represent the integral as a contour integral in the complex plane. We assume that $f(x)$ has an entire complex extension $f(z)$. Then $f(z)(\omega+z)^{-n}$ is analytic in the strip $[0,a]$ with a pole at $z=-\omega$ of order $n$. Then, the appropriate complex contour integral representation of the integral $S_{n}^{a}[f]$ is provided by equation \eqref{contourep1}. So that
\begin{equation}\label{bekbekk2}
	\int_0^a \frac{f(x)}{(\omega + x)^n}\,\mathrm{d}x  = \frac{1}{2\pi i} \int_{\mathrm{C}} \frac{f(z) \log z}{(\omega + z)^n} \mathrm{d}z -  \mathrm{Res} \left[\frac{f(z)}{(\omega+z)^n}\log z\right]_{z=-\omega} .
\end{equation}
    
We wish to manipulate the first term in equation \eqref{bekbekk2} to enable us later to identify the finite part integral. We introduce a trivial change in the first term with the replacement $\log z = (\log z -i\pi) + i\pi$. Then 
	\begin{eqnarray}
	\frac{1}{2\pi i} \int_{\mathrm{C}} \frac{f(z) \log z}{(\omega + z)^n} \mathrm{d}z &=& \frac{1}{2\pi i} \int_{\mathrm{C}} \frac{f(z)}{(\omega + z)^n} (\log z - i\pi) \mathrm{d}z \nonumber \\
	&& \hspace{18mm} +\,i \pi  \mathrm{Res}\left[\frac{f(z)}{(\omega + z)^n}\right]_{z=-\omega}
,	\label{sumsum}\end{eqnarray}
where we have used the fact that $f(z)$ is entire to arrive at the second term in the right hand side of the equation \eqref{sumsum}. Substituting equation \eqref{sumsum} back into equation \eqref{bekbekk2}, we obtain
\begin{eqnarray}
\int_0^a \frac{f(x)}{(\omega + x)^n}\mathrm{d}x &=& \frac{1}{2\pi i} \int_{\mathrm{C}} \frac{f(z)}{(\omega + z)^n} (\log z - i\pi) \mathrm{d}z \nonumber \\
&& \hspace{8mm}-\,\,\mathrm{Res} \left[\frac{f(z)}{(\omega + z)^n}(\log z - i \pi)\right]_{z=-\omega}\label{bokbok}
\end{eqnarray} 

Now we choose the contour $\mathrm{C}$ such that for a fixed $\omega$ and for all $z$ in $\mathrm{C}$ the following expansion converges absolutely,
	\begin{equation}\label{expansionn}
	\frac{1}{(\omega + z)^n} = \sum_{k=0}^{\infty} \binom{-n}{k} \frac{\omega^{k}}{z^{k + n}} .
	\end{equation}
	Absolute convergence is guaranteed provided $|\omega/z|<1$ for all $z$ in $\mathrm{C}$. This criterion can be simplified by deforming the contour of integration into a circle centered at the origin with radius $a$. Then the condition $|\omega/z|<1$ for all $z$ in $\mathrm{C}$ translates to the condition $\omega<a$, which is sufficient to impose for all paths of integration satisfying the condition. 
	
	We then introduce the expansion \eqref{expansionn} back into the first term in the right-hand side of equation \eqref{bokbok}. The absolute convergence of the binomial expansion along the contour of integration allows  us to interchange the order of integration and infinite summation in the first term.  Then
	\begin{eqnarray}\label{resultz}
	\int_0^a \frac{f(x)}{(\omega + x)^n} \mathrm{d}x &=& \sum_{k=0}^{\infty} \binom{-n}{k} \omega^{k} \frac{1}{2\pi i} \int_{\mathrm{C}} \frac{f(z)}{z^{k+n}} (\log z - i\pi) \mathrm{d}z \nonumber \\
	&& \hspace{8mm}- \mathrm{Res} \left[\frac{f(z)}{(\omega + z)^n}(\log z - i \pi)\right]_{z=-\omega}\label{bikbik}
	\end{eqnarray}
	Comparing the contour integrals in the first term with the contour integral representation of the finite part integrals for pole singularities given by equation \eqref{result1}, we find the integrals to be the finite parts of the divergent integrals $\int_0^a x^{-k-n} f(x)\, \mathrm{d}x$. 
	
	We can establish independently the absolute convergence of the infinite series in equation \eqref{bikbik}. We deform the contour of the contour integral representation of the finite part into a circle centered at the origin with the radius $a$. First, we have the bound
	\begin{eqnarray}
	\left|\bbint{0}{a} \frac{f(x)}{x^{k+n}}\mathrm{d}x\right| &=& \left|\frac{1}{2\pi i}\int_{\mathrm{C}} \frac{f(z)}{z^{k+n}} (\log z - i\pi) \mathrm{d}z\right|\nonumber \\
	&=& \left|\frac{1}{2\pi i} \int_0^{2\pi} \frac{f(a e^{i\theta})}{a^{k + n} e^{i(k+n)\theta}} (\ln a + (\theta - \pi)i) i a e^{i\theta} \mathrm{d}\theta\right|\nonumber \\
	&\leq & \frac{1}{a^{k+n - 1}} \frac{1}{2\pi} \int_0^{2\pi} \left|f(ae^{i\theta})(\ln a + (\theta - \pi)i)  \mathrm{d}\theta\right| \nonumber \\
	&\leq& \frac{1}{a^{k+n-1}} M(a)
	\end{eqnarray}
	where $M(a)$ is a finite positive constant independent of $k$.  Then we have the bound
	\begin{eqnarray}
		\left|\sum_{k=0}^{\infty} \binom{-n}{k} \omega^{ k} \bbint{0}{a} \frac{f(x)}{x^{k  +1}}\mathrm{d}x\right|&\leq& \sum_{k=0}^{\infty} \left|\binom{-n}{k}\right| \omega^{k} \left|\bbint{0}{a} \frac{f(x)}{x^{k  +1}}\mathrm{d}x\right| \nonumber \\
		&\leq& \frac{M(a)}{a^{n-1}}\sum_{k=0}^{\infty} \binom{n+k-1}{k} \frac{\omega^{ k}}{a^{k}} \nonumber \\
		&=& \frac{a M(a)}{(a - \omega)^n},
	\end{eqnarray}
	provided $a>\omega$. Then the infinite series converges absolutely.
	
Finally the residue term can be readily calculated, noting that we have a single pole at $z=-\omega$ of order $n$. The results are
\begin{equation}
\mathrm{Res}\left[\frac{f(z)}{(\omega + z)}\left(\log\,z-i\,\pi\right)\right]_{z=-\omega} =  f(-\omega) \ln\omega .
\end{equation}	
	\begin{eqnarray}
\mathrm{Res}\left[\frac{f(z)}{(\omega + z)^n}\left(\log\,z-i\,\pi\right)\right]_{z=-\omega}&=&\;\frac{1}{\left(n-1\right)!}f^{(n-1)}\!\left(-\omega\right)\, \ln\omega \nonumber \\
&& -\,\,\omega^{1-n}\sum_{k=0}^{n-2}\frac{f^{\left(k\right)}\left(-\omega\right)\,\omega^{k}}{k!\,\left(n-1-k\right)}, \;\;\; n=2, 3, \dots \label{main1sub}
\end{eqnarray}
Substituting all these back into the expansion \eqref{resultz} leads to the following result.
\begin{theorem}\label{theorem1}
Let the complex extension, $f(z)$, of $f(x)$ be entire. Then for all $n = 1, 2, 3, \dots$ and $0<\omega<a$, the following equality holds
	\begin{equation}\label{main1}
	\int_0^a \frac{f(x)}{(\omega + x)^n} \mathrm{d}x = \sum_{k=0}^{\infty} \binom{-n}{k} \omega^{ k} \bbint{0}{a} \frac{f(x)}{x^{k + n}}\mathrm{d}x + \Delta_{\mathrm{sc}}^{(n)}(\omega)
\end{equation}
where
\begin{equation}
\Delta_{\mathrm{sc}}^{(1)}(\omega)= - f(-\omega) \ln\omega .
\end{equation}	
	\begin{eqnarray}\label{singular_terrmm}
\Delta_{\mathrm{sc}}^{(n)}(\omega)&=&-\;\frac{1}{\left(n-1\right)!}f^{(n-1)}\!\left(-\omega\right)\, \ln\omega \nonumber \\
&& +\,\,\sum_{k=0}^{n-2}\frac{f^{\left(k\right)}\left(-\omega\right)}{k!\,\left(n-1-k\right) \; \omega^{n-k-1}}, \;\;\; n=2, 3, \dots \label{sing}
\end{eqnarray} 
\end{theorem}

\subsection{Behavior for small parameters}
If $f(z)$ is entire and analytic at the origin, which is our basic assumption on the function $f(z)$ in this paper, then we can always write it in the form $f(z)=z^m g(z)$, for some $m=0, 1, \dots $ and $g(z)=\sum_{k=0}^{\infty}d_k z^k$ with $d_0\neq 0$. When $m=0$, $f(z)$ does not vanish at the origin, $f(0)\neq 0$. On the other hand, when $m$ is a positive integer, $m=1, 2, \dots$, we have $f(0)=0$, so that $m$ is the order of zero of $f(z)$ at the origin. It will be our convention here to call the case $f(0)\neq 
0$ as $f(z)$ having a zero at the origin of order $m=0$. 

Now let us look into the dominant behavior of $S_n^a[f]$ as $\omega\rightarrow 0$. For $n=1$, the singular contribution dominates the finite part contributions as $\omega\rightarrow 0$ when $f(z)$ has a zero of order $m=0$ at the origin. Under this condition, we have the leading behavior
\begin{equation}
\int_0^a \frac{f(x)}{(\omega+x)}\, \mathrm{d}x \sim - f(0) \ln\omega,\;\;
\omega \rightarrow 0 .
\end{equation}
When $f(z)$ vanishes at the origin so that it has a zero of order $m>0$ there, 
we find that the leading contribution of $\Delta_{\mathrm{sc}}^{(1)}(\omega)$ is $\mathcal{O}(\omega^m \ln \omega)$, which vanishes as $\omega\rightarrow 0$. For this case the behavior near the origin is dominated by the leading finite part contribution,
\begin{equation}
\int_{0}^{a} \frac{f(x)}{(\omega+x)}\, \mathrm{d}x \sim \bbint{0}{a} \frac{f(x)}{x} \, \mathrm{d}x,\;\; \omega\rightarrow 0 .
\end{equation}
In fact the singular term will contribute only starting at the order $\mathcal{O}(\omega^m)$ of the naive contribution.

Similarly, for general $n =2,3,\dots$, the nature of the dominant contribution to the value of $S_{n}^{a}\left[f\right]$ depends chiefly on the order $m$ of the zero of $f\left(x\right)$ at the origin. First, we consider the case when $f$ has a zero at the origin of order $m = n-1$ so that we write $f\left(z\right) =  \sum_{j=0}^{\infty}d_j z^{j+n-1}$, with $d_0\neq 0$. The singular contribution \eqref{sing} assumes the form 
\begin{eqnarray}\label{dominant}
\Delta_{\mathrm{sc}}^{(n)}\left(\omega\right) &=& -\frac{\ln \omega}{\left(n-1\right)!}\,\sum_{j=0}^{\infty}\,\frac{\left(-1\right)^{j}\,\left(j+n-1\right)!\,d_{j}}{j!}\,\omega^{j}\\\nonumber
&&+\,\,\sum_{j=0}^{\infty}\,\sum_{k=0}^{n-1}\frac{\left(-1\right)^{j+n-k-1}\,\left(j+n-1\right)!\,d_j}{k!\,\left(n-1-k\right)\,\left(j+n-k-1\right)!}\,\omega^{j}
\end{eqnarray}
We see that in the limit as $\omega\to 0$, the first term of \eqref{dominant} provides the dominant contribution to the value of $S_{n}^{a}\left[f\right]$ due to the logarithmic factor so that
\begin{equation}
\int_0^a\frac{f(x)}{(\omega+x)^n}\, \mathrm{d}x \sim - d_0 \ln\omega , \;\; \omega\rightarrow 0 .
\end{equation}

When the order of the zero of $f$ at the origin is $m=0,1,\dots, (n-2)$ or $m=n-s$, for $s = 2,3,\dots,n$, the singular term provides the following dominant contribution
\begin{eqnarray}\label{dominant2}
\Delta_{\mathrm{sc}}^{(n)}\left(\omega\right) &=& -\frac{\ln \omega}{\left(n-1\right)!}\,\sum_{j=s-1}^{\infty}\,\frac{\left(j+n-s\right)!\,d_{j}}{\left(j-s+1\right)!}\,\left(-\omega\right)^{j-s+1}\\\nonumber
&&+\,\,\sum_{j=0}^{\infty}\,\sum_{k=0}^{n-s}\frac{\left(-1\right)^{j+n-k-s}\,\left(j+n-s\right)!\,d_j}{k!\,\left(n-1-k\right)\,\left(j+n-s-k\right)!}\,\omega^{j-s+1} \\\nonumber
&&+\,\,\sum_{k=n-s+1}^{n-2}\,\sum_{j=k+s-n}^{\infty}\frac{\left(-1\right)^{j+n-k-s}\,\left(j+n-s\right)!\,d_j}{k!\,\left(n-1-k\right)\,\left(j+n-s-k\right)!}\,\omega^{j-s+1}.
\end{eqnarray}
When $s=2$, the third term in the equation above is an empty sum over $k$, hence it vanishes. Furthermore, we see that the dominant contribution for all values of $s$ comes from the second term of \eqref{dominant2} which is of the order $\mathcal{O}\left(\omega^{-(s-1)}\right)$ so that
\begin{equation}
\int_0^a \frac{f(x)}{(\omega+x)^n} \, \mathrm{d}x \sim \frac{d_0}{\omega^{s-1}} \sum_{k=0}^{n-s} \frac{(-1)^{n-s-k}(n-s)!}{k!(n-1-k)(n-s-k)!} , \;\; \omega \rightarrow 0 .
\end{equation}

Similarly, when the order of the zero of $f$ is $m=n, (n+1), (n+2), \dots$ or $m=n+r$ for $r=0,1,2,\dots$, the singular contribution \eqref{singular_terrmm} assumes the form
\begin{eqnarray}
\Delta_{\mathrm{sc}}^{(n)}\left(\omega\right) &=& -\frac{\ln \omega}{\left(n-1\right)!}\,\sum_{j=0}^{\infty}\,\frac{\left(j+n+r\right)!\,d_{j}}{\left(j+r+1\right)!}\,\left(-\omega\right)^{j+r+1}\\\nonumber
&&+\,\,\sum_{j=0}^{\infty}\,\sum_{k=0}^{n-2}\frac{\left(-1\right)^{j+n-k+r}\,\left(j+n+r\right)!\,d_j}{k!\,\left(n-1-k\right)\,\left(j+n-k+r\right)!}\,\omega^{j+r+1} ,
\end{eqnarray}
which has the leading contribution $\mathcal{O}(\omega^{r+1}\ln\omega)$. The singular contribution then vanishes as $\omega\rightarrow 0$ and ceases to be the dominant term. It is now dominated by the leading term of the naive contribution so that
\begin{equation}
\int_0^a \frac{f(x)}{(\omega+x)^n}\, \mathrm{d}x \sim \bbint{0}{a}\frac{f(x)}{x^n} \, \mathrm{d}x, \;\; \omega\rightarrow 0 .
\end{equation}

Since the order of zero of $f(z)$ is equal to or greater than $n$, the leading finite part integral is a convergent integral. In fact all terms up to $\mathcal{O}(\omega^r)$ in the naive term of equation \eqref{main1} are convergent integrals whose values are equal to the finite part integral (see Remark 1 in Section-\ref{finitepartintegration}). Clearly, the singular contributions start to appear from the first divergent term in the naive expansion, i.e. for $k=r+1$. 

\subsection{Example} The Gauss hypergeometric function ${_2F_1}$ admits the following Euler integral representation \cite[p558, eq.15.3.1]{abramowitz},
\begin{equation}\label{hypergeom}
\pFq{2}{1}{\sigma,a}{b}{-z}=\frac{\Gamma(b)}{\Gamma(a) \Gamma(b-a)} \int_0^1 \frac{x^{a-1} (1-x)^{b-a-1}}{(1+z x)^{\sigma}} \mathrm{d}x , \;\; b>a>0 .
\end{equation}
This is an example of an incomplete Stieltjes transform. This representation can be brought into the form amenable to treatment by Theorem-\ref{theorem1} for specific values of the parameters $a$, $b$ and $\sigma$. Let $a=r, b=s, \sigma=n$ be all positive integers, with $s-r>1$ to ensure local integrability of $(1-x)^{b-a-1}$ at $x=1$. Moreover, we let $z=\zeta>0$ and just simply effect an analytic continuation later for complex values when desired. Then the representation \eqref{hypergeom} assumes the form
\begin{equation}\label{hypergeom_eg}
\pFq{2}{1}{n,r}{s}{-\zeta}=\frac{(s-1)!}{(r-1)!\,(s-r-1)!\, \zeta^n} \int_0^1 \frac{x^{r-1} (1-x)^{s-r-1}}{(\zeta^{-1}+x)^{n}} \mathrm{d}x  .
\end{equation}
In this form, the theorem can now be applied with $f(x)=x^{r-1} (1-x)^{s-r-1}$ and $\omega=\zeta^{-1}$. The complexification of $f(x)$ is the entire, in fact polynomial, function $f(z)=z^{r-1} (1-z)^{s-r-1}$. Since the theorem provides an expansion in the neighborhood of $\omega = 0$, the resulting expansion of the hypergeometric function is an expansion about infinity or for arbitrarily large $z$. Moreover, this expansion is convergent so that we will have a convergent asymptotic expansion of the hypergeometric function.

When $r=1$, $f(z)$ does not vanish at the origin; and when $r > 1$, $f(z)$ has a zero there of order $r-1$. We consider the case $n\geq r$ for which the singular contribution dominates the naive contribution. Then applying Theorem \ref{theorem1} to the integral in \eqref{hypergeom_eg}, we obtain
\begin{align}\label{integarl_part}
\int_{0}^{1}\frac{x^{r-1}\left(1-x\right)^{s-r-1}}{\left(\zeta^{-1}+x\right)^{n}}\mathrm{d}x = \sum_{k=0}^{\infty}{{-n}\choose{k}}\zeta^{-k}\bbint{0}{1}\frac{x^{r-1}\left(1-x\right)^{s-r-1}}{x^{k+n}}\mathrm{d}x+\Delta_{\mathrm{sc}}^{(n)}\left(\zeta^{-1}\right)
\end{align}
Expanding the integrand and then using our result above for polynomials \eqref{integer_polynomial}, the finite part integral is obtained to be
\begin{align}\label{finite_partt}
\bbint{0}{1}\frac{x^{r-1}\left(1-x\right)^{s-r-1}}{x^{k+n}}\mathrm{d}x = \sum_{l=0}^{s-r-1}\frac{\left(s-r-1\right)!\,\left(-1\right)^{l}}{l!\,\left(s-r-1-l\right)!\left(l+r-k-n\right)}.
\end{align}
On the other hand, evaluating the indicated differentiation in equation \eqref{sing} yields the singular contributions, 
\begin{align}\label{missing_termss}
\Delta_{sc}^{(n)} \left(\zeta^{-1}\right) = \frac{\left(-1\right)^{r-n}\,\mathrm{ln}\,\zeta}{\zeta^{s-n-1}\,\left(\zeta+1\right)^{r+1-s}}\sum_{k=0}^{n-1}\frac{\left(r-1\right)!\,\left(s-r-1\right)!\,\left(\zeta+1\right)^{-k}}{k!\,\left(n-1-k\right)!\,\left(r+k-n\right)!\,\left(s-r-k-1\right)!}\\\nonumber
+\frac{\zeta^{n+1-s}\,\left(-1\right)^{r-1}}{\left(\zeta+1\right)^{r+1-s}}\sum_{k=0}^{n-2}\sum_{l=0}^{k}\frac{\left(r-1\right)!\,\left(s-r-1\right)!\,\left(\zeta+1\right)^{-l}\,\left(-1\right)^{k}}{l!\,\left(k-l\right)!\,\left(r+l-k-1\right)!\,\left(s-r-l-1\right)!\,\left(n-1-k\right)} .
\end{align}
For positive integers $n$, $r$, $s$ such that $r+1<s<n+1$, the summation in the first term of the right-hand side of equation \eqref{missing_termss} vanishes for $0\leq k<n-r$ due to the negative argument of the second factorial in the denominator while for $k\geq n-r$, the third factorial vanishes as well. Since $r\geq 1$ the sum is equal to zero in  the entire summation range. Then we have the particular case where the logarithmic term in the singular contribution in equation \eqref{singular_terrmm} of Theorem-\ref{theorem1} vanishes. 

Hence, upon substitution of equations \eqref{finite_partt} and \eqref{missing_termss} back into equation \eqref{integarl_part}, we obtain the following series representation of the Gauss hypergeometric function \eqref{hypergeom_eg} 
\begin{eqnarray}\nonumber\label{opem}
\pFq{2}{1}{n,r}{s}{-\zeta} &=&\frac{\left(s-1\right)!}{\left(n-1\right)!\,\left(r-1\right)!}\, \sum_{k=0}^{\infty}\frac{\left(n+k-1\right)!\,\left(-1\right)^{k}}{k!}\,\frac{a_k}{\zeta^{k+n}}\\
&&+\,\frac{\left(-1\right)^{r-1}\,\left(s-1\right)!}{\zeta^{s-1}\,\left(\zeta+1\right)^{r+1-s}}\,\sum_{m=0}^{n-2}\frac{c_m}{m! (s-r-m-1)!}\frac{1}{(1+\zeta)^m}\,\label{bekbek}
\end{eqnarray}
where
\begin{equation}
a_k = \sum_{l=0}^{s-r-1}\frac{\left(-1\right)^{l}}{l!\,\left(s-r-1-l\right)!\,\left(l+r-k-n\right)}
\end{equation}
and
\begin{align}
c_m = \sum_{j=m}^{n-2}\frac{(-1)^j}{(n-1-j)(j-m)!(r+m-j-1)!}.
\end{align}
for   $\zeta>1$ and $r+1<s<n+1$, with $s, r, n$ positive integers. Since the series representation of the Gauss hypergeometric function for positive powers of its argument converges at most in the entire unit circle, equation \eqref{bekbek} serves as an extension of the canonical infinite series representation of $\,_2F_1(a,b;c;z)$ for $z=-\zeta<-1$ under the stated conditions.

We can rewrite equation \eqref{bekbek} by expressing the first term in the right-hand side in terms of hypergeometric function itself so that we obtain the relation,
\begin{eqnarray}\nonumber\label{gaussone}
\pFq{2}{1}{n,r}{s}{-\zeta} &=&\frac{(-1)^{s+r}\,(s-1)!\,(n-s)!}{(r-1)!\,(n-r)!\,\zeta^{n}}\pFq{2}{1}{n,n-s+1}{n-r+1}{-\frac{1}{\zeta}}\\
&&+\,\frac{\left(-1\right)^{r-1}\,\left(s-1\right)!}{\zeta^{s-1}\,\left(\zeta+1\right)^{r+1-s}}\,\sum_{m=0}^{n-2}\frac{c_m}{m! (s-r-m-1)!}\frac{1}{(1+\zeta)^m}.
\end{eqnarray}
Invoking the analytic properties of $\,_2F_1$, this relation can be extended beyond $\zeta>1$ into the entire complex plane. Furthermore, we obtain the following asymptotic behavior,
\begin{eqnarray}\nonumber\label{gauss1}
\pFq{2}{1}{n,r}{s}{-\zeta} &=& \frac{\left(s-1\right)!}{\left(r-1\right)!\,\zeta^n}\,\left(a_0+\mathcal{O}\left(\frac{1}{\zeta}\right)\right) \nonumber \\
&+&\,\,\frac{\left(-1\right)^{r-1}\,\left(s-1\right)!}{\zeta^{s-1}\,\left(\zeta+1\right)^{r+1-s}(s-r-1)!}\,\left(c_0 + \mathcal{O}\left(\frac{1}{1+\zeta}\right)\right), \,\;\; \zeta\rightarrow\infty.
\end{eqnarray}

The correctness of equation \eqref{bekbek} can be independently verified by an explicit evaluation of its right hand side and comparing the result with the tabulated values of its left hand side. Mathematica\textsuperscript \textregistered tabulates values of $\,_2F_1$ for positive integer values of the parameters.  
For instance when $r=2$, $s=4$, and $n=5$, Mathematica\textsuperscript \textregistered returns,
\begin{equation}
\pFq{2}{1}{5,2}{4}{-\zeta} = \frac{\zeta +2}{2 (\zeta +1)^3} .
\end{equation}
This value is reproduced by equation \eqref{bekbek} when the given parameters are substituted back and the resulting expression appropriately simplified. While specific values of $\,_2F_1$ can be automatically generated by algebra systems (for positive integer parameters), the series representation given by equation \eqref{bekbek}, together with its attendant corollaries, equations \eqref{gaussone} and \eqref{gauss1}, are not tabulated in \cite{prudnikov,NIST,abramowitz,ghauss}.

\subsection{Example}
The Kummer function of the second kind $U\left(a,b,z\right)$ has the integral representation \cite[p326, eq.13.4.4]{NIST},
\begin{align}\label{Kummer's_second}
U(a,b,z)=\frac{1}{\Gamma(a)} \int_0^{\infty} e^{-z t} t^{a-1}(1+t)^{b-a-1} \, \mathrm{d}t, \;\; \mathrm{Re}\,a>0, |\mathrm{arg}\, z|<\frac{\pi}{2} .
\end{align}
A new series representation of the Kummer function can be obtained from this representation by means of Theorem-\ref{theorem1} for the specialized values of the parameter $z=\omega>0$, $b=a+1-n$ for $n=1,2,\dots$ and integer $a=s\geq 1$. For these values of the parameters, equation \eqref{Kummer's_second} reduces to the form 
\begin{equation}
U\left(s,s+1-n,\omega\right) = \frac{1}{\left(s-1\right)!\,\omega^{s-n}}\,\int_{0}^{\infty}\frac{e^{-x}\,x^{s-1}}{\left(\omega+x\right)^{n}}\,\mathrm{d}x,
\end{equation}
where a change in variable $x\rightarrow \omega x$ has been performed to obtain the integral. 

We identify the integral as a generalized Stieltjes transform of integral order $n$ for the function $f(x)=x^{s-1} e^{-x}$. The complexified function $f(z)=z^{s-1} e^{-z}$ is entire so that Theorem \ref{theorem1} applies and it yields \begin{align}\label{integ_kummer}
\int_{0}^{\infty}\frac{e^{-x}\,x^{s-1}}{\left(\omega+x\right)^{n}}\,\mathrm{d}x = \sum_{k=0}^{\infty}{{-n}\choose{k}}\,\omega^{k}\bbint{0}{\infty}\frac{e^{-x}\,x^{s-1}}{x^{k+n}}\,\mathrm{d}x+\Delta_{\mathrm{sc}}^{(n)}\left(\omega\right).
\end{align}
Now $f(z)$ does not vanish at the origin for $s=1$ and has a zero of order $s-1$ there for all positive integer $s>1$. Let us consider the two cases $n\geq s$ and $1\leq n<s$. In the former, the singular contribution dominates the naive contribution; while in the latter, the singular contribution is subdominant to the naive contribution.

First, we consider the case $n\geq s$. Under this condition, all integrals in the naive term by term integration in equation \eqref{integ_kummer} are divergent. The corresponding finite part integrals are specialized values of equation \eqref{expo1} and are given by 
\begin{equation}
\bbint{0}{\infty}\frac{e^{-x}}{x^{k+n+1-s}}\,\mathrm{d}x = \frac{\left(-1\right)^{k+n-s}}{\left(k+n-s\right)!}\,\psi\left(k+n+1-s\right),\;\; n\geq s, \;\;\; k=0, 1, \cdots .
\end{equation}
The singular contributions are computed to be
\begin{align}\nonumber\label{piyto}
\Delta_{\mathrm{sc}}^{(n)}\left(\omega\right) &= \left(-1\right)^{n+s+1}\,\ln\omega\,e^{\omega}\,\left(s-1\right)!\,\omega^{s-1}\sum_{j=0}^{n-1}\frac{\omega^{-j}}{j!\left(n-1-j\right)!\,\left(s-j-1\right)!}\\
&+\left(-1\right)^{s-1}\,e^{\omega}\,\left(s-1\right)!\,\omega^{s-n}\sum_{k=0}^{n-2}\sum_{j=0}^{k}\frac{\omega^{k-j}\,\left(-1\right)^{k}}{j!\,\left(k-j\right)!\,\left(s-j-1\right)!\,\left(n-1-k\right)}
\end{align}

Substituting all the contributions back into the integral in \eqref{integ_kummer} and after performing a resummation of the second term in equation \eqref{piyto}, the Kummer function of the second kind assumes the following exact representation
\begin{eqnarray}\label{joji}\nonumber
U\left(s,s+1-n,\omega\right) = \frac{\left(-1\right)^{n-s}\,\omega^{n-s}}{\left(s-1\right)!\,\left(n-1\right)!}\sum_{k=0}^{\infty}\frac{\left(n+k-1\right)!\,\psi\left(k+n+1-s\right)}{\left(k+n-s\right)!}\frac{\omega^{k}}{k!}\\
\hspace{-20mm}+\,\left(-1\right)^{n+s+1}\,e^{\omega}\,\mathrm{ln}\,\omega\,\sum_{j=0}^{n-1}\frac{\omega^{n-1-j}}{j!\,\left(n-1-j\right)!\,\left(s-j-1\right)!}\\\nonumber
+\,\left(-1\right)^{s-1}\,e^{\omega}\sum_{m=0}^{n-2}\frac{\omega^{m}}{m!}d_{m}\hspace{4.1cm}
\end{eqnarray}
where 
\begin{equation}\label{fity}
d_m = \sum_{l=0}^{n-2-m}\frac{(-1)^{l+m}}{l!(s-l-1)!\,(n-1-l-m)}
\end{equation}
for all $n=1, 2, 3, \dots$, $s=1,\dots (n-1), n$  and $\omega>0$, where an empty sum is equal to zero. The representation given by equation \eqref{joji} is untabulated in \cite{prudnikov,NIST,abramowitz,trikummer}. 

We can obtain an independent verification of the representation given by equation \eqref{joji} by comparing its special values with tabulated values of the Kummer function. We have, in particular, from Reference \cite{a_2} the value 
\begin{eqnarray}\nonumber
U\left(2,-4,\omega\right) = \frac{1}{720}\left(e^{\omega}(\omega+6)(\mathrm{Chi}(\omega)-\mathrm{Shi}(\omega))\omega^{5}\right.\hspace{3cm}
\\
\hspace{3cm}+(\omega(\omega(\omega(\omega+5)-4)+6)-12)\omega+24\big)\label{biik}
\end{eqnarray}
where $\mathrm{Chi(\omega)}$ and $\mathrm{Shi(\omega)}$ are the hyperbolic cosine and hyperbolic sine integrals, respectively. The Kummer function $U(2,-4,\omega)$ corresponds to $s=2$ and $n=7$ of equation \eqref{joji}. Substituting these parameters back into equation \eqref{joji} and performing simplification with the aid of Mathematica\textsuperscript\textregistered reproduces equation \eqref{biik}.

Now from equation \eqref{joji}, we obtain the following asymptotic behavior of the Kummer function,
\begin{eqnarray}\label{hitr0}\nonumber
U\left(s,s+1-n,\omega\right) &=& \frac{(-1)^{n-s}\,\psi(n+1-s)\,\omega^{n-s}}{(s-1)!(n-s)!}(1+\mathcal{O}(\omega))\\\nonumber
&&+\,(-1)^{n+s+1}\,e^{\omega}\,\ln\omega\left(\frac{1}{(n-1)!\,(s-n)!}+\mathcal{O}(\omega)\right)\\
&&+\,(-1)^{s-1}\,e^{\omega}\left(d_0+\mathcal{O}\left(\omega\right)\right), \;\;\; \omega\rightarrow 0 .
\end{eqnarray}
where $d_0$ is given as 
\begin{equation}
d_0= \sum_{l=0}^{n-2}\frac{(-1)^{l}}{l!\,(s-l-1)!(n-1-l)} = \frac{(-1)^n (\psi (s)-\psi(-n+s+1))}{\Gamma (n) \Gamma (-n+s+1)},
\end{equation}
in which $\psi(z)$ is the digamma function. The dominant terms originate from the second and third terms of the right hand side of equation \eqref{joji} which are the singular contribution given in equation \eqref{piyto}. Since $s\leq n$, the factor $1/(s-n)!$ in the second term vanishes except when $s=n$.  We can compare this with the known asymptotic behavior of $U(a,b,z)$, in particular \cite[p508, eq.13.5.9]{abramowitz}
\begin{equation}\label{pitn}
U(a,1,z)= -\frac{1}{\Gamma(a)}\left[\ln z+\psi(a)+2\gamma\right]+\mathcal{O}(|z\ln z|), \;\; z\rightarrow 0.
\end{equation}
The $b=1$ case corresponds to $s=n$ in \eqref{hitr0}; under this condition, equation \eqref{hitr0} reduces to
\begin{equation}\label{ponit}
U(s,1,\omega) = -\frac{2\gamma}{(s-1)!}\left(1+\mathcal{O}(\omega)\right)-e^{\omega}\ln\omega\left(\frac{1}{(s-1)!}+\mathcal{O}(\omega)\right)-\frac{e^{\omega}\psi(s)}{(s-1)!}.
\end{equation} 
Equation \eqref{ponit} reduces to equation \eqref{pitn} with the replacement $e^{\omega}=1+\omega + \mathcal{O}(\omega^2)$ in \eqref{ponit}. For positive integers $a=s=1, 2, \dots$, equation \eqref{ponit} contains more asymptotic information than equation \eqref{pitn}.

We now consider the case $n<s$. Under this condition, the first $(s-n-1)$ integrals in the naive term of equation \eqref{integ_kummer} are convergent and the remaining integrals are all divergent. We can then split the infinite series in \eqref{integ_kummer} to collect together the convergent and the divergent contributions as follows 
\begin{eqnarray}\label{case2_kummer}
\int_{0}^{\infty}\frac{e^{-x}\,x^{s-1}}{\left(\omega+x\right)^{n}}\,\mathrm{d}x &=& \sum_{k=0}^{s-n-1}{{-n}\choose{k}}\,\omega^{k}\bbint{0}{\infty}x^{s-n-k-1}\,e^{-x}\mathrm{d}x\\\nonumber
&&+ \sum_{k=s-n}^{\infty}{{-n}\choose{k}}\,\omega^{k}\bbint{0}{\infty}\frac{e^{-x}}{x^{k+n-s+1}}\,\mathrm{d}x + \Delta_{\mathrm{sc}}^{(n)}\left(\omega\right)
\end{eqnarray}
The integrals appearing in the first sum of the right hand side of the equation above are convergent integrals. They are evaluated as regular (Riemann) integrals,
\begin{equation}
\bbint{0}{\infty}x^{s-n-k-1}\,e^{-x}\mathrm{d}x = \int_{0}^{\infty}x^{s-n-k-1}\,e^{-x}\mathrm{d}x = \left(s-n-k-1\right)! .
\end{equation}

The rest of the terms in the right hand side of equation \eqref{case2_kummer} are evaluated as finite part integrals in the same manner as in the previous case so that we obtain the following exact expansion of the Kummer function of the second kind
\begin{eqnarray}\nonumber\label{hihoo}
U\left(s,s+1-n,\omega\right) &=& \frac{\omega^{n-s}}{\left(s-1\right)!\,\left(n-1\right)!}\sum_{k=0}^{s-n-1}\left(n+k-1\right)!\,\left(s-n-k-1\right)!\,\frac{\left(-\omega\right)^{k}}{k!}\\\nonumber
&&+\frac{\left(-1\right)^{n-s}\,\omega^{n-s}}{\left(s-1\right)!\,\left(n-1\right)!}\sum_{k=s-n}^{\infty}\frac{\left(n+k-1\right)!\,\psi\left(k+n+1-s\right)}{\left(k+n-s\right)!}\frac{\omega^{k}}{k!}\\\nonumber
&&+\left(-1\right)^{n+s+1}\,e^{\omega}\,\mathrm{ln}\,\omega\,\sum_{j=0}^{n-1}\frac{\omega^{n-1-j}}{j!\,\left(n-1-j\right)!\,\left(s-j-1\right)!}\\
&&+\,\left(-1\right)^{s-1}\,e^{\omega}\sum_{m=0}^{n-2}\frac{\omega^{m}}{m!}d_{m}
\end{eqnarray}
where $d_m$ is given by equation \eqref{fity}, $s=2,3,4,\dots$, while $n=1,2,\dots,(s-1)$, and $\omega>0$; any empty sum is equal to zero. This representation, too, is untabulated in \cite{prudnikov,NIST,abramowitz,trikummer}.

Again we can independently validate our representation \eqref{hihoo} by comparing its special values with tabulated values of $U(a,b,z)$.  We have in \cite{a_5} the result
\begin{equation}\label{momo}
U\left(5,2,\omega\right) = \frac{(\omega+3)(\omega(\omega+8)+2)+e^{\omega}(\omega(\omega+6)^{2}+24)(\mathrm{Chi}(\omega)-\mathrm{Shi}(\omega))}{144\omega}
\end{equation}
The parameters of $U(5,2,z)$ correspond to $s=5$ and $n=4$ in equation \eqref{hihoo}. Substituting these parameters back into \eqref{hihoo} leads to the same expression \eqref{momo}.

Now from equation \eqref{hihoo} we obtain the following asymptotic behavior of the Kummer function,
\begin{eqnarray}\nonumber\label{hitr2}
U\left(s,s+1-n,\omega\right) &=& \frac{(s-n-1)!}{(s-1)!\,\omega^{s-n}}\left(1+\mathcal{O(\omega)}\right) + \frac{(-1)^{n-s+1}\,\gamma}{(s-n)!\,(n-1)!}(1+\mathcal{O}(\omega)) \\
&&+\,\frac{(-1)^{n+s+1}\,e^{\omega}\,\ln\omega}{(n-1)!\,(s-n)!}\left(1+\mathcal{O(\omega)}\right)+\,(-1)^{s-1}\,e^{\omega}\left(d_0+\mathcal{O}\left(\omega\right)\right)
\end{eqnarray}
as $\omega\to 0$ under the stated conditions.  We compare this with the asymptotic behavior tabulated in \cite[p508, eq.13.5.6 and eq.13.5.7]{abramowitz} as $z\rightarrow 0$
\begin{eqnarray}\label{wa1}
U(a,b,z) &=& \frac{\Gamma(b-1)}{\Gamma(a)}z^{1-b}+\mathcal{O}\left(|z|^{\mathrm{Re}(b)-2}\right),\,\,\mathrm{Re}(b)\geq 2,b\neq 2\\\label{yuko}
&=&\frac{\Gamma(b-1)}{\Gamma(a)}z^{1-b}+\mathcal{O}\left(\ln z\right),\,\,b=2 .\label{wa2}
\end{eqnarray}
Equation \eqref{wa1} corresponds to the dominant first term of \eqref{hitr2} for $s\neq (n-1)$; and equation \eqref{wa2} for $s=n-1$. Equations \eqref{wa1} and \eqref{yuko} miss the information coming from the subdominant terms of \eqref{hitr2}. 

\section{Generalized Stieltjes transform of integral orders of entire functions with branch point at the origin }\label{integralbranch}

This time let us consider the finite part integration of the incomplete generalized Stieltjes transform of integral order $n$ of the function $h(x)=x^{-\nu} f(x)$, where $f(x)$ has an entire complex extension $f(z)$ and $0<\nu<1$,
\begin{align}\label{incomplte}
S_{n}^a\left[h\right] = \int_{0}^{a}\frac{x^{-\nu}\,f\left(x\right)}{\left(\omega+x\right)^{n}}\,\mathrm{d}x
\end{align}
To perform finite-part integration, we  represent the right hand side of equation \eqref{incomplte}, as a contour integral using Lemma-\ref{lemma0}, with the contour enclosing the pole $z=-\omega$. 
\begin{eqnarray}\label{eq:lemaa1}
\int_{0}^{a}\,\frac{x^{-\nu}\,f\left(x\right)}{(\omega+z)^n}\,\mathrm{d}x&=&\frac{1}{e^{-2\,\pi\,\nu\,i}-1}\,\int_{C}\,\frac{f\left(z\right)}{z^{\nu}(\omega+z)^n}\,\mathrm{d}z \nonumber \\
&&\hspace{24mm}-\frac{2\,\pi\,i}{e^{-2\,\pi\,\nu\,i}-1}\;\mathrm{Res}\left[\frac{z^{-\nu}\,f\left(z\right)}{(\omega+z)^n}\right]_{z=-\omega}
\end{eqnarray}
We then proceed in the same manner as in proving Theorem-\ref{prop1}. The result is given by the following Theorem.

\begin{theorem}\label{theorem2}
Let the complex extension, $f\left(z\right)$, of $f\left(x\right)$ be entire. Then for all $n=1,2,3,\dots$, $0<\omega<a$, and $0<\nu<1$, the following equality holds
\begin{equation}\label{branch2}
\int_{0}^{a}\,\frac{x^{-\nu}f\left(x\right)}{\left(\omega+x\right)^{n}}\,\mathrm{d}x=\sum_{k=0}^{\infty}\,{{-n}\choose{k}}\,\omega^{k}\,\bbint{0}{a}\,\frac{f\left(x\right)}{x^{n+k+\nu}}\,\mathrm{d}x+\Delta_{\mathrm{sc}}^{(n)}\left(\omega\right)
\end{equation}
where
\begin{equation}\label{singular_cont}
\Delta_{\mathrm{sc}}^{(n)}\left(\omega\right)=\frac{\pi}{\sin\left(\pi\,\nu\right)\,\omega^{\nu}}\,\sum_{k=0}^{n-1}\,\frac{f^{(n-1-k)}\left(-\omega\right)}{k!\,\left(n-1-k\right)!}\,\frac{\left(\nu\right)_{k}}{\omega^{k}} .
\end{equation}
\end{theorem}

\subsection{Behavior for small parameters} We now obtain the explicit dependence of the nature of the dominant contribution to the value of $S_{n}^{a}\left[f\right]$ on the order of the zero of $f$ at the origin. For $f$ with a zero of order $m=0, 1, \dots$, we write $f\left(z\right) = \sum_{j=0}^{\infty}d_{j}z^{j+m}$, with $d_0 \neq 0$. The singular contribution \eqref{singular_cont} assumes the form
\begin{align}\label{singular_explicit}
\Delta_{\mathrm{sc}}^{(n)}\left(\omega\right) = \frac{\pi}{\sin\left(\pi\,\nu\right)\,\omega^{\nu}}\,\sum_{j=0}^{\infty}\,\sum_{k=0}^{n-1}\frac{\left(-1\right)^{k}\,d_j\,\left(j+m\right)!\,\left(\nu\right)_k\,\left(-\omega\right)^{j+m-n+1}}{k!\,\left(n-1-k\right)!\,\left(j+m-n+k+1\right)!}.
\end{align}
For $m=0,1,\dots,n-1$, the singular term provides the dominant contribution as $\omega\to 0$. In particular, when $m=n-1$, the leading term of \eqref{singular_explicit}, that is for $j=0$, becomes
\begin{align}
\int_0^a \frac{x^{-\nu} f(x)}{(\omega+x)^n}\,\mathrm{d}x \sim \frac{\pi\,d_0\,\left(n-1\right)!\,}{\sin\left(\pi\,\nu\right)\,\omega^{\nu}}\,\sum_{k=0}^{n-1}\frac{(-1)^k \,\left(\nu\right)_k}{\left(k!\right)^{2}\,\left(n-1-k\right)!\,},\qquad\omega\to 0
\end{align}

On the other hand, when $m = n+r$, for $r=0,1,2,\dots$, the singular contribution merely provides a leading order correction term to the dominant contribution coming from the naive term. In particular, when $r=0$, that is $m=n$, the leading term of the singular contribution \eqref{singular_explicit} assumes the form 
\begin{align}
 \Delta_{\mathrm{sc}}^{(n)}(\omega) \sim -\frac{\pi\,d_0\,n!\,\omega^{1-\nu}}{\sin\left(\pi\,\nu\right)}\,\sum_{k=0}^{n-1}\frac{(-1)^k\,\left(\nu\right)_k}{k!\,\left(n-1-k\right)!\,\left(k+1\right)!},\qquad\omega\to 0
\end{align}
In this case, the leading singular contribution is of the order $\mathcal{O}\left(\omega^{1-\nu}\right)$ which is dominated by the leading naive term. Then the Stieltjes transform $S_{n}^{a}\left[f\right]$ has the leading behavior
\begin{equation}
\int_0^a \frac{x^{-\nu} f(x)}{(\omega+x)^n}\, \mathrm{d}x \sim \bbint{0}{a} \frac{x^{-\nu} f(x)}{x^n}\, \mathrm{d}x , \;\; \omega\rightarrow 0 .
\end{equation}
As in the previous case, this leading contribution is a convergent integral and correction terms coming from the singular contribution appear from the first divergent term arising from term by term integration.

\subsection{Example} Let us consider again the Gauss hypergeometric function for the following set of parameters
\begin{equation}\label{pako}
\pFq{2}{1}{n,1-\mu}{s-\mu+2}{-z} = \frac{\Gamma(s-\mu+2)}{\Gamma(1-\mu) \Gamma(s+1) z^n} \int_0^1 \frac{x^{-\mu} (1-x)^s}{(z^{-1}+x)^n} \mathrm{d}x
\end{equation}
for $n,s=1, 2, 3\dots$, and $0<\mu<1$. Without loss of generality, we assume that $z=\zeta>0$ so that $\omega=\zeta^{-1}>0$. Applying Theorem-\ref{theorem2} on the integral, we obtain
\begin{equation}\label{peks}
\int_0^1 \frac{(1-x)^s}{x^{\mu}\, (\zeta^{-1} + x)^n}\,\mathrm{d}x = \sum_{k=0}^{\infty} {-n \choose k}\, \frac{1}{\zeta^{k}}\,\bbint{0}{1} \frac{(1-x)^s}{x^{n+k+\mu}}\,\mathrm{d}x + 
\Delta_{\mathrm{sc}}^{(n)}(\zeta^{-1}), \;\; \zeta>1 ,
\end{equation}
where the finite part integrals, obtained using equation \eqref{polynomialbranch}, are given by
\begin{equation}
\bbint{0}{1} \frac{(1-x)^s}{x^{n+k+\mu}} \mathrm{d}x = \sum_{l=0}^s \frac{(-1)^l s!}{l!\,(s-l)! (l-n-k-\mu +1)!},
\end{equation}
and the singular contributions are given by
\begin{equation}
\Delta_{\mathrm{sc}}^{(n)}(\zeta^{-1})= (-1)^{n-1}\frac{\pi s! }{\sin(\pi \mu)} \sum_{k=0}^{n-1} \frac{(-1)^k\,\left(\mu\right)_k\,(1 + \zeta^{-1})^{s-n+k+1}}{k!\,(s-n+k+1)!\,(n-k-1)!}\,\zeta^{k+\mu},
\end{equation}
Notice that the singular contributions dominate the terms coming from the finite part integrals for $\zeta\rightarrow \infty$. 

Substituting all contributions back into the integral \eqref{peks} and then to equation \eqref{pako}, the Gauss hypergeometric function assumes the following exact representation for the specified parameters
\begin{eqnarray}\label{youi} 
\pFq{2}{1}{n,1-\mu}{s-\mu+2}{-\zeta} = \frac{\Gamma(s-\mu+2)}{\Gamma(1-\mu)\, (n-1)!\, \zeta   ^n} \sum_{k=0}^{\infty} \,\frac{\left(-1\right)^{k}\,(n+k-1)!}{k!\,}\,\frac{b_k}{\zeta^k} \\\nonumber
+\frac{\left(-1\right)^{n+1}\,\Gamma(s-\mu+2)(1+\zeta)^{s+1}}{\zeta^{s-\mu+1}}\sum_{k=0}^{n-1}\frac{(-1)^{k}\,\Gamma(\mu+k)}{k!\,(s-n+k+1)!\,(n-k-1)!\,(1+\zeta)^{n-k}} , 
\end{eqnarray}
where
\begin{equation}
b_k=\sum_{l=0}^s \frac{(-1)^{l}}{l!\,(s-l)!\,(l-n-k-\mu+1)} = \frac{\Gamma\left(1-\mu-n-k\right)}{\Gamma(s-\mu+2-n-k)}
\end{equation}
for all $\zeta>1$, $n,s=1, 2, 3, \dots$ and $0<\mu<1$.

The result can be written in terms of hypergeometric function so that we arrive at the following relation 
\begin{eqnarray}\nonumber\label{gausstwo}
\pFq{2}{1}{n,1-\mu}{s-\mu+2}{-\zeta} = \frac{\Gamma(1-\mu-n)\,\Gamma(s-\mu+2)}{\Gamma(1-\mu)\,\Gamma(s-\mu+2-n)\,\zeta^{n}}\pFq{2}{1}{n,n+\mu-s-1}{n+\mu}{-\frac{1}{\zeta}}\\
+\frac{(-1)^{n-1}\,\Gamma(\mu)\,\Gamma(s-\mu+2)}{\zeta^{s-\mu+1}\,(1+\zeta)^{n-s-1}\,\Gamma(s-n+2)\,\Gamma(n)}\pFq{2}{1}{\mu,1-n}{s-n+2}{1+\zeta} .
\end{eqnarray}
Also from the same representation \eqref{youi}, we obtain the following asymptotic behavior
\begin{eqnarray}\label{gauss2}\nonumber
\pFq{2}{1}{n,1-\mu}{s-\mu+2}{-\zeta}=\frac{\Gamma(s-\mu+2)}{\Gamma(1-\mu)\,\zeta^n}\left(b_0 + \mathcal{O}\left(\frac{1}{\zeta}\right)\right)\hspace{3.5cm}\\
+\frac{\Gamma(s-\mu+2)\Gamma(\mu+n-1)(1+\zeta)^{s}}{s!\,(n-1)!\,\zeta^{s-\mu+1}}\left(1+\mathcal{O}\left(\frac{1}{1+\zeta}\right)\right)
\end{eqnarray}
as $\zeta\to \infty$. 

We can again verify the correctness of the representation given by equation \eqref{youi} by comparing it with tabulated values of the Gauss hypergeometric function. For example, we have the known value \cite[p474,\#99]
{prudnikov} 
\begin{equation}
\pFq{2}{1}{2,\frac{1}{2}}{\frac{5}{2}}{-\zeta} = \frac{3 \left(\sqrt{\zeta }+(\zeta -1) \tan ^{-1}\left(\sqrt{\zeta }\right)\right)}{4 \zeta ^{3/2}} .\label{eke}
\end{equation}
The left hand side corresponds to the parameters $n=2$, $\mu=1/2$, and $s=1$ in the representation \eqref{youi}. Substituting these values back into the right hand side of equation \eqref{youi} reproduces equation  \eqref{eke}.

For the case of positive integer values of the parameters of the hypergeometric function $\,_2F_1(a,b;c;z)$, Mathematica\textsuperscript \textregistered and Maple\textsuperscript \textregistered  tabulate specific values produced by the representation given by equation \eqref{opem}. However, for non-integral values of the parameters $b$ and $c$, in general both algebra systems return the hypergeometric function unevaluated. But representation \eqref{youi} may allow evaluation of $\,_2F_1(a,b;c;z)$ in terms of elementary functions. For instance, for $n=2$, $\mu=1/3$, $s=1$, and $\zeta>1$, representation \eqref{youi} yields the value
\begin{eqnarray}\nonumber
\pFq{2}{1}{2,\frac{2}{3}}{\frac{8}{3}}{-\zeta} = \frac{10}{9\,\zeta}\left(1+\frac{\zeta^{\frac{1}{3}}}{3}\left(\frac{2}{\zeta}-1\right)\left(\mathrm{ln}\left[1+\frac{1}{\zeta^{\frac{1}{3}}}\right]-\frac{1}{2}\mathrm{ln}\left[1-\frac{1}{\zeta^{\frac{1}{3}}}+\frac{1}{\zeta^{\frac{2}{3}}}\right]\right.\right.\\
\left.\left.+\sqrt{3}\,\tan^{-1}\left[\frac{\sqrt{3}}{2}\frac{\frac{1}{\zeta^{\frac{1}{3}}}}{1-\frac{1}{2\,\zeta^{\frac{1}{3}}}}\right]\right)\right) + \frac{20}{81}\frac{\sqrt{3}\,\pi(\zeta-2)}{\zeta^{\frac{5}{3}}};\label{baba1}
\end{eqnarray}
and for $n=3$, $s=3$, $\mu=1/2$,
\begin{equation}
\pFq{2}{1}{3,\frac{1}{2}}{\frac{9}{2}}{-\zeta} =\frac{35 \left((\zeta -3) \sqrt{\zeta } (3 \zeta +5)+3 (\zeta +1)
   ((\zeta -2) \zeta +5) \tan ^{-1}\left(\sqrt{\zeta
   }\right)\right)}{128 \zeta ^{7/2}}.\label{baba2}
\end{equation}
For these examples, both Mathematica\textsuperscript \textregistered and Maple\textsuperscript \textregistered return $\,_2F_1(a,b;c;z)$ unevaluated. One can reproduce these results by substituting the parameters back into equation \eqref{youi} and then performing simplification using Maple\textsuperscript \textregistered for equation \eqref{baba1} and Mathematica\textsuperscript \textregistered for \eqref{baba2}. The general results \eqref{youi} and the corollaries derived from it, as well as the special cases \eqref{baba1} and \eqref{baba2} are untabulated in \cite{prudnikov,NIST,abramowitz,ghauss}.

\subsection{Example}
We again obtain an exact series expansion of the Kummer function of the second kind for the following specific parameters $z=\omega>0$, $0<a<1$, $b=a-n+1$ for $n=1,2,\dots$ by way of Theorem-\ref{theorem2}. Thus, we rewrite the integral \eqref{Kummer's_second} by making the substitution $x=\omega\,t$ so that the function assumes the form of a Stieltjes transform
\begin{align}\label{whittaker}
U\left(a,a-n+1,\omega\right) = \frac{\omega^{n-a}}{\Gamma\left(a\right)}\,\int_{0}^{\infty}\,\frac{e^{-x}}{x^{1-a}\,\left(\omega+x\right)^n}\,\mathrm{d}x\qquad n=1,2,\dots
\end{align}

Applying Theorem-\ref{theorem2} to the integral above, we have
\begin{align}\label{apply_theorem}
\int_{0}^{\infty}\,\frac{e^{-x}}{x^{1-a}\,\left(\omega+x\right)^{n}}\,\mathrm{d}x &= \sum_{k=0}^{\infty}\,{{-n}\choose{k}}\,\omega^{k}\,\bbint{0}{\infty}\,\frac{e^{-x}}{x^{n+k+1-a}}\,\mathrm{d}x + \Delta_{\mathrm{sc}}^{(n)}\left(\omega\right).
\end{align}
The finite part integral in the RHS of the equation above is given by equation \eqref{expofinite}, in particular,
\begin{align}
\bbint{0}{\infty}\,\frac{e^{-x}}{x^{n+k+1-a}}\,\mathrm{d}x = \frac{\left(-1\right)^{n+k}\,\pi}{\Gamma\left(n+k+1-a\right)\,\sin\left(\pi\,a\right)}
\end{align}
On the other hand, the singular contributions are computed to be
\begin{eqnarray}\label{hitr}
\Delta_{\mathrm{sc}}^{(n)}\left(\omega\right) 
=\frac{\left(-1\right)^{n+1}\,\pi\,e^{\omega}\,\omega^{a-1}}{\sin\left(\pi\,a\right)\,\Gamma\left(1-a\right)}\,\sum_{k=0}^{n-1}\,\frac{\Gamma\left(k+1-a\right)}{k!\,\left(n-1-k\right)!\,\left(-\omega\right)^{k}}
\end{eqnarray}
Substituting these results to \eqref{whittaker} and simplifying, we obtain a series representation for the Kummer function of the second kind
\begin{eqnarray}\nonumber\label{pino}
U\left(a,a-n+1,\omega\right) &=& \frac{\left(-1\right)^{n}\,\Gamma\left(1-a\right)\,\omega^{n-a}}{\left(n-1\right)!}\,\sum_{k=0}^{\infty}\,\frac{\left(n+k-1\right)!}{\Gamma\left(n+k+1-a\right)}\,\frac{\omega^{k}}{k!}\\
&&+\left(-1\right)^{n+1}\,e^{\omega}\,\omega^{n-1}\,\sum_{k=0}^{n-1}\,\frac{\Gamma\left(k+1-a\right)}{k!\,\left(n-1-k\right)!\,\left(-\omega\right)^{k}} .
\end{eqnarray}
for all $0<a<1$, $n=1, 2, 3, \dots$ and $\omega>0$. 

Several results corresponding to  specific cases of equation \eqref{pino} exist and can be used to verify its validity. For instance, when $a=1/2$ and $n=3$ equation \eqref{pino} reduces to the following known result given in \cite{a_half},
\begin{equation}
U\left(\frac{1}{2},-\frac{3}{2},\omega\right) = \frac{1}{8}\left(2\sqrt{\omega}(3-2\omega)+e^{\omega}\sqrt{\pi}(4\omega(\omega-1)+3)\mathrm{erfc}\left(\sqrt{\omega}\right)\right)
\end{equation}
where other tabulated results exist corresponding to $a=1/2$ and positive integer $n$ up to $n=7$ but an exact master representation such as equation \eqref{pino} of comparable simplicity from which these special cases could be generated is not, to our knowledge, available. Furthermore, we found minimal treatment on the asymptotic behavior of the Kummer function of the second kind for the specific set of parameters considered here in the $\omega\to0$ limit. Only the relatively simple cases where $U(a,b,\omega)$ reduces to a polynomial have been considered in \cite{NIST}. While in \cite{prudnikov}, no tabulated expansions about $\omega=0$ were given. 

On the other hand, the  results in equations \eqref{hitr0}, \eqref{hitr2}, and \eqref{hitr3} may serve to supplement the tabulated cases found in \cite{asymp} which is by no means exhaustive. We may also generate special cases for which no representation is given by leading computer algebra systems such as Mathematica\textsuperscript \textregistered and Maple\textsuperscript \textregistered  . For instance when $a=1/7$, $n=2$ and $\omega>0$, equation \eqref{pino} reduces to
\begin{equation}
U\left(\frac{1}{7},-\frac{6}{7},\omega\right) = \frac{1}{6} \left(7 \omega ^{13/7}-e^{\omega } (7 \omega -6) \Gamma
   \left(\frac{13}{7},\omega \right)\right)
\end{equation}
where $\Gamma(\alpha,\omega)$ is the incomplete gamma function; and for $a=1/6$, $n=4$, and $\omega>0$
\begin{eqnarray}\nonumber
U\left(\frac{1}{6},-\frac{17}{6},\omega\right) = \frac{\omega ^{23/6}}{5610} \small\left(-e^{\omega } (18 \omega  (6 \omega  (2 \omega -5)+55)-935) E_{-\frac{17}{6}}(\omega )\right.\\
+72 \omega  (3 \omega
   +1)+582\Big)\hspace{2cm}
\end{eqnarray}
where $E_{b}(\omega )$ is the exponential integral function. With the aid of the representation \eqref{pino}, an exhaustive list of new representations for families of special cases can be generated for $U(a,b,\omega)$.

From equation \eqref{pino}, the following limiting behavior can be directly extracted 
\begin{equation}\label{hitr3}
U(a,a-n+1,\omega) =  \frac{(-1)^{n}\Gamma(1-a)\omega^{n-a}}{{\Gamma(n+1-a)}}(1+\mathcal{O}(\omega))+e^{\omega}\,\frac{\Gamma(n-a)}{(n-1)!}\left(1+\mathcal{O}(\omega)\right),
\end{equation}
as $\omega\to 0$ where the dominant second term is provided by the singular contribution \eqref{hitr}. We compare equation \eqref{hitr3} with the expansion for $U(a,b,z)$ in \cite[p508, eq.13.5.12]{abramowitz} as $|z|\to 0$
\begin{equation}
U(a,b,z) = \frac{\Gamma(1-b)}{\Gamma(1+a-b)}+\mathcal{O}(|z|),\,\,\mathrm{Re}(b)\leq 0, b\neq 0
\end{equation}
This limiting form corresponds to the dominant second term in equation \eqref{hitr3} with the exponential factor replaced by its leading small-$\omega$ approximation which is unity. Consequently our result is a numerically improved expansion.

\section{Some Physical applications}\label{application}
Finite-part integration finds application in problems where the physical property that is the object of computation assumes a Stieltjes representation. For example, in laminar and turbulent flows the effective diffusivity tensor assumes the representation
\begin{equation}\label{diffusion}
\kappa_{\mathrm{eff}}=\kappa \left[1 + \int_{-\infty}^{\infty} \frac{{Pe}^2}{1+{Pe}^2  \tau^2 }\mu(\mathrm{d}\tau)\right]
\end{equation}
where $\mu(\mathrm{d}t)$ is a positive matrix-valued measure on $(-\infty,\infty)$ and $Pe$ is the Peclet number \cite{avellada}. The integral involved in equation \eqref{diffusion} is a generalized Stietljes transform. In application, the large Peclet number behavior of the effective diffusitivity is desired; that is, one is interested in obtaining the asymptotic expansion of equation \eqref{diffusion} as $Pe\rightarrow\infty$. An attempt to obtain the expansion by expanding $(1+Pe^2 \tau^2)^{-1}$ about $Pe=\infty$ and then integrating term by term leads to the infinite series
\begin{equation}
1 + \int_{-\infty}^{\infty}\frac{\mu(\mathrm{d}\tau)} {\tau^2} - \frac{1}{Pe^2}\int_{-\infty}^{\infty}\frac{\mu(\mathrm{d}\tau)} {\tau^4} + \frac{1}{Pe^4}\int_{-\infty}^{\infty}\frac{\mu(\mathrm{d}\tau)} {\tau^6} - \dots  
\end{equation}
Each term, except the first term, is a divergent integral so that the attempted solution fails. This problem has been addressed by following a circuitous approach involving two steps. 
The first step is to obtain the asymptotic expansion (which is in general divergent) of the Stieltjes transform for small Peclet number; this involves calculating the moments $\int_{-\infty}^{\infty} \tau^m \, \mu(\mathrm{d}\tau)$ for positive integers $m$. The second step is to sum the divergent asymptotic expansion by means of Pade approximants; the large Peclet number behavior is obtained by an extension of the Pade approximant to large $Pe$'s. The result of the last step must be considered closely from a numerical stand point to assess how well does the Pade approximants derived from the asymptotic series of the Stieltjes integral continue to give meaningful values even in the domain of large $Pe$. Such assessment must be made in contrast to more direct and suitable alternatives one of which we shall propose here.

Under certain conditions, finite-part integration obviates the need to go through the two-step process to obtain the desired high Peclet number behavior. If the measure is given by $\mu(\mathrm{d}\tau)=g(\tau) \mathrm{d}\tau$, we can cast the Stieltjes transform into a form that is amenable to finite-part integration as we have developed above. Splitting the integration among the positive and negative real axis and changing variable in the negative side, the transform assumes the form 
\begin{equation}\label{kwak}
\int_{-\infty}^{\infty} \frac{Pe^2\,g(\tau)}{1+ Pe^2 \tau^2}\,\mathrm{d}\tau = \int_0^{\infty} \frac{Pe^2\,g(\tau)}{1+ Pe^2 \tau^2}\,\mathrm{d}\tau + \int_0^{\infty} \frac{Pe^2\,g(-\tau)}{1+ Pe^2 \tau^2}\,\mathrm{d}\tau .
\end{equation}
Letting $\omega=1/Pe$, the two integrals are special cases of the Stieltjes transform
\begin{equation}\label{kwek}
\mathrm{S}[f] = \int_{0}^{a}\frac{f(x)}{\omega^{2}+x^{2}}\mathrm{d}x, 0<a\leq\infty,
\end{equation}
so that the problem of evaluating equation \eqref{kwak} for large Peclet number $Pe$ reduces to evaluating equation \eqref{kwek} for small values of $\omega$. 

The same analysis can be applied to obtain an asymptotic expansion of the Stieltjes integral representation of the Green-Kubo formula for the diffusion coefficient along the direction of an arbitrary unit vector $e$,
\begin{equation}\label{pinpy}
D^{e}=\frac{1}{\gamma}\vert\vert\widehat{V}^{e}_{N}\vert\vert^{2}_{\mathcal{H}}+2\gamma \int_{0}^{\infty}\frac{\mathrm{d}\mu_e(\lambda)}{\gamma^{2}+\lambda^{2}},
\end{equation}
for an arbitrary vector $\widehat{V}^{e}_{N}$ where $\mathrm{d}\mu_{e} = \left<\mathrm{d}P(\lambda)\widehat{V}^{e}_{N^{\perp}},\widehat{V}^{e}_{N^{\perp}}\right>$ as well as the antisymmetric part of the diffusion tensor 
\begin{equation}
A_{ij} = \int_{\mathbb{R}}\frac{\lambda\mathrm{d}\mu_{ij}(\lambda)}{\lambda^{2}+\gamma^{2}},
\end{equation}
both in the weak noise limit $\gamma\to 0$. An advantage of these representations is that an application of finite-part integration to obtain a suitable expansion of the Stieltjes integral  will provide improved asymptotic information on the limiting behavior of these quantities. For instance, the diffusion coefficient has the limiting behavior, 
$\mathrm{lim}_{\gamma\to 0}\,\gamma D^{e} = \vert\vert\widehat{V}^{e}_{N}\vert\vert^{2}_{\mathcal{H}}$
(Proposition 3.3 in \cite{pavliotis}), derived from a different representation, so that $D^e$ at least behaves asymptotically like
$D^{e}=\mathcal{O}\left(\gamma^{-1}\right)$ as $\gamma\to 0$. Finite part integration allows obtaining a more accurate asymptotic estimate, if not an exact result.

We now proceed in performing finite-part integration of equation \eqref{kwek} assuming that the real-valued $f(x)$ possesses an entire complex extension $f(z)$. This is accomplished by an application of Lemma \ref{lemma1}. Let $C$ be a circular path centered at the origin with radius $a>\omega$. Then by Lemma-\ref{lemma1} we arrive at the representation 
\begin{equation}\label{tint}
\int_{0}^{a}\frac{f(x)}{\omega^{2}+x^{2}}\mathrm{d}x = \frac{1}{2\pi i }\int_{C}\frac{f(z)\,\log z}{\omega^{2}+z^{2}}\mathrm{d}z - \sum_{k}\mathrm{Res}\left[\frac{f(z)\log z}{\omega^{2}+z^{2}}\right]
\end{equation}
where the residues are due to the poles at $z= \pm\omega i$. A zero can be added to the first term so that it assumes the form
\begin{equation}\label{ngit}
\frac{1}{2\pi i}\int_{C}\frac{f(z)\,\log z}{\omega^{2}+z^{2}}\mathrm{d}z = \frac{1}{2\pi i}\int_{C}\frac{f(z)\,(\log z-\pi i)}{\omega^{2}+z^{2}}\mathrm{d}z + \frac{\pi i}{2\pi i}\int_{C}\frac{f(z)}{\omega^{2}+z^{2}}\mathrm{d}z
\end{equation}
A term by term integration can then be performed on the first term of the right-hand side of equation \eqref{ngit} by writing 
\begin{equation}\label{expansion2}
\frac{1}{\omega^{2}+z^{2}} = \frac{1}{z^{2}\left(1+\frac{\omega^{2}}{z^{2}}\right)} = \frac{1}{z^{2}}\sum_{k=0}^{\infty}(-1)^{k}\frac{\omega^{2k}}{z^{2k}}
\end{equation}
and interchanging the operations of summation and integration. The interchange is justified because expansion \eqref{expansion2} converges uniformly along the contour of integration. 

Since $f(z)$ is entire, the second term in equation \eqref{ngit} is just the sum of residues due to the simple poles at $z=\pm i \omega$ so that 
\begin{equation}
\frac{\pi i}{2\pi i}\int_{C}\frac{f(z)}{\omega^{2}+z^{2}}\mathrm{d}z = \sum_{k}\mathrm{Res}\left[\frac{f(z)}{\omega^{2}+z^{2}}\right] .
\end{equation}
Hence equation \eqref{tint} can be written as 
\begin{equation}\label{penge}
\int_{0}^{a}\frac{f(x)}{\omega^{2}+x^{2}}\mathrm{d}x = \sum_{k=0}^{\infty}(-1)^{k}\frac{\omega^{2k}}{2\pi i}\int_{C}\frac{f(z)(\log z-\pi i)}{z^{2k+2}}\mathrm{d}z - \sum_{k}\mathrm{Res}\left[\frac{f(z)(\log z - \pi i)}{\omega^{2}+z^{2}}\right]
\end{equation}
The residue evaluates to
\begin{eqnarray}\nonumber
\sum_{k}\mathrm{Res}\left[\frac{f(z)(\log z - \pi i)}{\omega^{2}+z^{2}}\right] 
&=& -\frac{\pi}{2}\mathrm{Re}f(\omega i) + \frac{\ln \omega}{\omega}\mathrm{Im}f(\omega i)
\end{eqnarray}
We then make use of equation \eqref{result1} to identify the relevant finite part in the first term of the right hand side of equation \eqref{penge} so that we obtain the result
\begin{equation}\label{finalresult}
\int_{0}^{a}\frac{f(x)}{\omega^{2}+x^{2}}\mathrm{d}x = \sum_{k=0}^{\infty}(-1)^{k}\omega^{2k}\bbint{0}{a}\frac{f(x)}{x^{2k+2}}\mathrm{d}x +\frac{\pi}{2\omega}\mathrm{Re}f(\omega i) - \frac{\ln \omega}{\omega}\mathrm{Im}f(\omega i) .
\end{equation}

Equation \eqref{finalresult} gives the full expansion of the Stieltjes transform about $\omega=0$. The dominant behavior as $\omega\rightarrow 0$ is controlled by the real and the imaginary parts of $f(i\omega)$. For the case of calculating the effective diffusivity, this result saves ample numerical resources  by doing away with calculating mere approximants to the Stieltjes integral. Furthermore, applying the result in equation \eqref{finalresult} to the Green-Kubo formula for the diffusion coefficient \eqref{pinpy}, we see that we obtain an additional non-negligible logarithmic term in the weak noise limit $\gamma\to 0$ when it happens that the imaginary part of $f(i\omega)$ does not vanish, for example, when $f(x)= e^{-x}$. The same treatment can be done in evaluating the antisymmetric part of the diffusion tensor for which no small $\gamma$ behavior was explored in the rather abstract treatment given in \cite{pavliotis}. These demonstrate several advantages of finite part integration, not least translating the relatively formidable problem of obtaining a suitable expansion for the effective diffusivity corresponding to $Pe\rightarrow\infty$, the diffusion coefficient, and the antisymmetric part of the diffusion tensor in the  weak noise limit to mere calculations of low order residues. We shall explore the full physical ramifications of these results in substantial detail elsewhere. 

\section{Conclusion}\label{conclusion}
In this paper, we have evaluated the incomplete generalized Stieltjes transform of integral order by finite-part integration in a form that allows us to extract the asymptotic behavior of the transform for small values of the parameter. We have seen once more that an attempt to evaluate the Stieltjes integral by expanding the integrand and performing term by term integration lead to missing terms. By finite-part integration, we were able to recover the missing terms which are contributions coming from the poles and the branch points of the integrand in the complex plane. When the function under transformation does not vanish or it has a zero at the origin whose order does not sufficiently exceed the order of the Stieltjes transformation, the missing terms are the dominant terms for arbitrarily small values of the parameter of the transformation. Moreover, we have seen that finite part integration is not only a means of obtaining the leading asymptotic behavior of the Stieltjes integral but also is a potent tool in the exact evaluation of functions that admit Stieltjes transform representations. This allows obtaining new representations of  the special functions of mathematical physics that may lead to their evaluations in terms of simpler functions and in uncovering relationships among themselves. Most important is that finite part integration offers a new tool in investigating physical phenomena whose observables or relevant physical quantities appear as Stieltjes integrals.  This in turn, leads to the possibility of gaining unprecedented insights and perhaps novel understanding of the processes from which the physical quantities arise.

\section*{Acknowledgment}
The authors acknowledge the Office of the Chancellor of the University of the Philippines Diliman, through the Office of the Vice Chancellor for Research and Development, for funding support through the Outright Research Grant 171711 PNSE.

\end{document}